\documentclass[acmtog]{acmart}
\pdfoutput=1
\acmSubmissionID{643}

\usepackage{mathtools}

\usepackage{booktabs} % For formal tables

\usepackage{graphicx}

\usepackage{subfig}

\begin{document}

\setcopyright{rightsretained} 
\acmJournal{TOG}
\acmYear{2018}\acmVolume{37}\acmNumber{4}\acmArticle{68}\acmMonth{8} \acmDOI{10.1145/3197517.3201401}

%Conference
%\acmConference[SIGGRAPH '18]{SIGGRAPH '18: Special Interest Group on Computer Graphics and Interactive Techniques Conference Technical Paper Proceedings }{August 2018}{Vancouver, British Columbia Canada}
%\acmYear{1997}
%\copyrightyear{2016}

\title{Deep Appearance Models for Face Rendering}

\author{Stephen Lombardi}
\affiliation{%
  \institution{Facebook Reality Labs}
  \city{Pittsburgh}
  \state{PA}}
\email{stephen.lombardi@fb.com}

\author{Jason Saragih}
\affiliation{%
  \institution{Facebook Reality Labs}
  \city{Pittsburgh}
  \state{PA}}
\email{jason.saragih@fb.com}

\author{Tomas Simon}
\affiliation{%
  \institution{Facebook Reality Labs}
  \city{Pittsburgh}
  \state{PA}}
\email{tomas.simon@fb.com}

\author{Yaser Sheikh}
\affiliation{%
  \institution{Facebook Reality Labs}
  \city{Pittsburgh}
  \state{PA}}
\email{yaser.sheikh@fb.com}

%\affiliation{%
%  \institution{Facebook Reality Labs}
%  \city{Pittsburgh}
%  \state{PA}}

\renewcommand{\shortauthors}{Lombardi et al.}

\begin{teaserfigure}
    \centering
    \includegraphics[trim={0 4mm 1mm 0},width=1.0\textwidth]{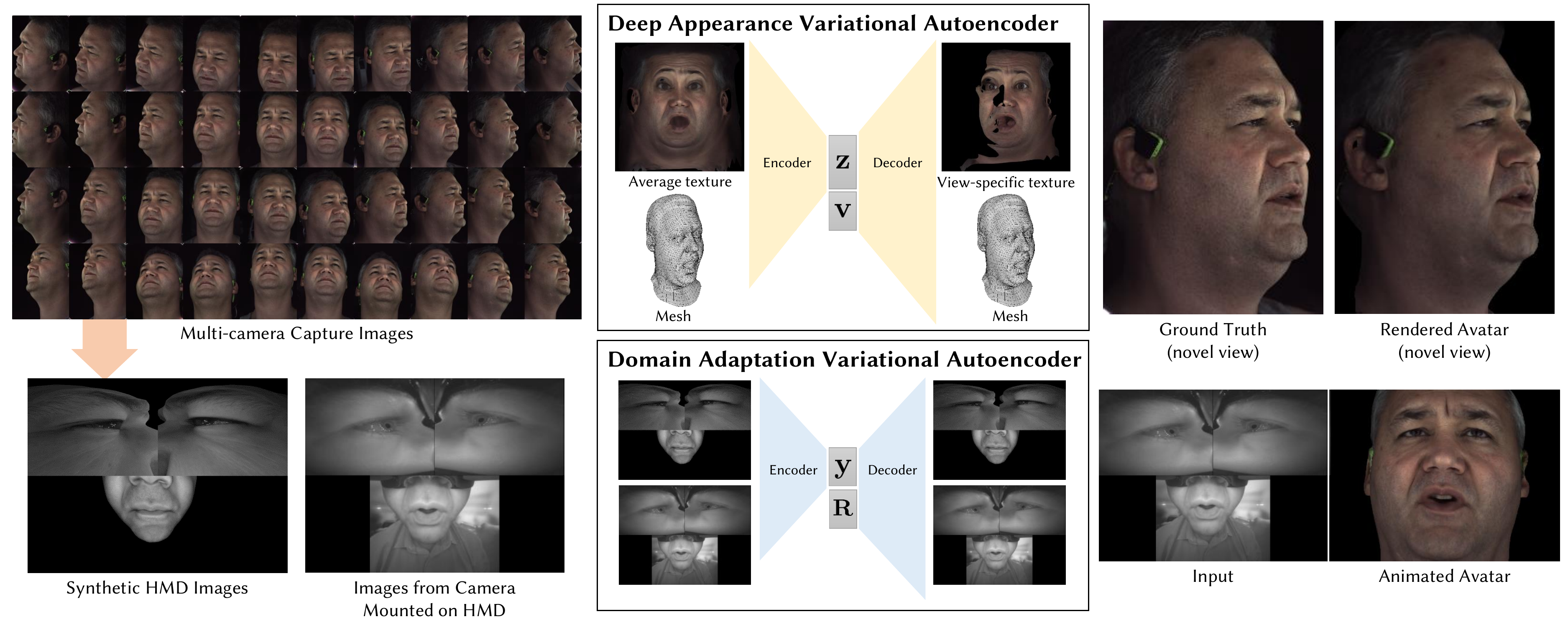}
    \caption{Our model jointly encodes and decodes geometry and view-dependent appearance into a latent code $\mathbf{z}$, from data captured from a multi-camera rig, enabling highly realistic data-driven facial rendering. We use this rich data to drive our avatars from cameras mounted on a head-mounted display (HMD). We do this by creating synthetic HMD images through image-based rendering, and using another variational autoencoder to learn a common representation $\mathbf{y}$ of real and synthetic HMD images. We then regress from $\mathbf{y}$ to the latent rendering code $\mathbf{z}$ and decode into mesh and texture to render. Our method enables high-fidelity social interaction in virtual reality. }
    \label{fig:teaser}
\end{teaserfigure}

\begin{abstract}
We introduce a deep appearance model for rendering the human face. Inspired by Active Appearance Models, we develop a data-driven rendering pipeline that learns a joint representation of facial geometry and appearance from a multiview capture setup. Vertex positions and view-specific textures are modeled using a deep variational autoencoder that captures complex nonlinear effects while producing a smooth and compact latent representation. View-specific texture enables the modeling of view-dependent effects such as specularity. In addition, it can also correct for imperfect geometry stemming from biased or low resolution estimates. This is a significant departure from the traditional graphics pipeline, which requires highly accurate geometry as well as all elements of the shading model to achieve realism through physically-inspired light transport. Acquiring such a high level of accuracy is difficult in practice, especially for complex and intricate parts of the face, such as eyelashes and the oral cavity. These are handled naturally by our approach, which does not rely on precise estimates of geometry. Instead, the shading model accommodates deficiencies in geometry though the flexibility afforded by the neural network employed. At inference time, we condition the decoding network on the viewpoint of the camera in order to generate the appropriate texture for rendering. The resulting system can be implemented simply using existing rendering engines through dynamic textures with flat lighting. This representation, together with a novel unsupervised technique for mapping images to facial states, results in a system that is naturally suited to real-time interactive settings such as Virtual Reality (VR). 

%In addition, we present a semi-supervised technique for relating our representation to images captured under the viewing conditions commonly encountered in Virtual Reality (VR) applications, where existing approaches for estimating facial motion are inadequate. Together, these approaches enable compelling high-fidelity facial animation suitable for social interactions in VR. 

%The latent space of facial appeance and geometry in deep appearance models is not human interpretable. This presents a challenge when animating the model either by hand or automatically through video-based techniques. To enable interoperability of our model with existing animation pipelines, 

%we build a mapping function that relates the deep appearance model's latent space to a human interpretable action-unit set, commonly used in 3D face tracking systems. 
%In addition, we present a technique for relating our representation to images captured under extreeme viewing conditions, commonly encountered in Virtual Reality (VR) applications, where existing approaches for estimating facial motion are inadequatte. 
%We provide qualitative results showing that our method produces compelling realism, quantitative results evaluating our model, and we devise a method to drive our avatars with cameras mounted on a head-mounted display.
\end{abstract}




\maketitle

\section{Introduction}

% Background: virtual reality, immersion, and social presence
With the advent of modern Virtual Reality (VR) headsets, there is a need for improved real-time computer graphics models for enhanced immersion. Traditional computer graphics techniques are capable of creating highly realistic renders of static scenes but at high computational cost. These models also depend on physically accurate estimates of geometry and shading model components. When high precision estimates are difficult to acquire, degradation in perceptual quality can be pronounced and difficult to control and anticipate. Finally, photo-realistic rendering of dynamic scenes in real time remains a challenge.

%This is especially the case for the human face, to which humans are extremely perceptually sensitive.

% Problem: social interactions in VR
Rendering the human face is particularly challenging. Humans are social animals that have evolved to express and read emotions in each other's faces~\cite{ekman1980face}. As a result, humans are extremely sensitive to rendering artifacts, which gives rise to the well-known uncanny valley in photo-realistic facial rendering and animation. The source of these artifacts can be attributed to deficiencies, both in the rendering model as well as its parameters. This problem is exacerbated in real-time applications, such as in VR, where limited compute budget necessitates approximations in the light-transport models used to render the face. Common examples here include: material properties of the face that are complex and time-consuming to simulate; fine geometric structures, such as eyelashes, pores, and vellus hair that are difficult to model; and subtle dynamics from motion, particularly during speech. Although compelling synthetic renderings of faces have been demonstrated in the movie industry, these models require significant manual cleanup~\cite{Lewis2014PracticeAT}, and often a frame-by-frame adjustment is employed to alleviate deficiencies of the underlying model. This manual process makes them unsuitable for real-time applications.

%but also critical for immersive social experiences. In order to provide a virtual experience with all the richness and depth of a face-to-face interaction, we must capture and display all the nuances of the social signals between participants.

% Crisis
%There are many reasons why achieving photorealistic human face rendering in real-time has been difficult: the material properties of the face are complex and time-consuming to simulate; the human face contains fine geometric structures, such as eyelashes, pores, and vellus hair that are difficult to model; the face exhibits subtle dynamics during motion and in particular for speech. Additionally, human perception is more acute on faces because we are social animals that have evolved to express and read emotions on each other's faces \cite{}.

The primary challenge posed by the traditional graphics pipeline stems form its physics-inspired model of light transport. For a specific known object class, like a human face, it is possible to circumvent this generic representation, and instead directly address the problem of generating images of the object using image statistics from a set of examples. This is the approach taken in Active Appearance Models (AAMs)~\cite{Cootes:TPAMI:2001}, which synthesize images of the face by employing a joint linear model of both texture and geometry (typically a small set of 2D points), learned from a collection of face images. Through a process coined analysis-by-synthesis, AAMs were originally designed to infer the underlying geometry of the face in unseen images by optimizing the parameters of the linear model so that the synthesized result best matches the target image. Thus, the most important property of AAMs is their ability to accurately synthesize face images while being sufficiently constrained to only generate plausible faces. Although AAMs have been largely supplanted by direct regression methods for geometry registration problems~\cite{Xiong-2013-7701,kazemi2014one}, their synthesis capabilities are of interest due to two factors. First, they can generate highly realistic face images from sparse 2D geometry. This is in contrast to physics-inspired face rendering that requires accurate, high-resolution geometry and material properties. Second, the perceptual quality of AAM synthesis degrades gracefully. This factor is instrumental in a number of perceptual experiments that rely on the uncanny-valley being overcome~\cite{cite:Zoe,cite:AAMFace}. 

% Active Appearance Models
%Active Appearance Models (AAMs)~\cite{Cootes:TPAMI:2001,Edwards:ICFGR:1998} have been used in computer vision to register human faces in images by using a joint linear subspace of geometry (often a small set of 2D points) and texture. These models have largely been supplanted for computer vision tasks by modern methods that directly regress from images to  facial landmarks or facial expression. AAMs perform analysis by synthesis, finding the best setting of the latent parameters that causes the ``decoded'' image to best match the input image. This synthesized image is typically thrown out as the interest is primarily in the latent parameters. Our primary goal, on the other hand, is to use morphable models for photorealistic image synthesis.

% High level idea of our model
Inspired by the AAM, in this work we develop a data-driven representation of the face that jointly models variations in sparse \emph{3D} geometry and \emph{view-dependent} texture. Departing from the linear models employed in AAMs, we use a deep conditional variational autoencoder~\cite{Kingma:CoRR:2013} (CVAE) to learn the latent embedding of facial states and decode them into rendering elements (geometry and texture). A key component of our approach is the view-dependent texture synthesis, which can account for limitations posed by sparse geometry as well as complex nonlinear shading effects such as specularity. We explicitly factor out viewpoint from the latent representation, as it is extrinsic to the facial performance, and instead condition the decoder on the direction from which the model is viewed. To learn a deep appearance model of the face, we constructed a multiview capture apparatus with 40 cameras pointed at the front hemisphere of the face. A coarse geometric template is registered to all frames in a performance and it, along with the unwarped textures of each camera, constitute the data used to train the CVAE. We investigated the role of geometric precision and viewpoint granularity and found that, although direct deep image generation techniques such as those of~\citet{Radford:CoRR:2015}, \citet{Hou:WACV:2017}, or~\citet{Kulkarni:NIPS:2015} can faithfully reconstruct data, extrapolation to novel viewpoints is greatly enhanced by separating the coarse graphics warp (via a triangle mesh) from appearance, confirming the basic premise of the parameterization. 

Our proposed approach is well suited for visualizing faces in VR since; a) sparse geometry and dynamic texture are basic building blocks of modern real-time rendering engines, and b) modern VR hardware necessarily estimates viewing directions in real-time. However, in order to enable interaction using deep appearance models in VR, the user's facial state needs to be estimated from a collection of sensors that typically have extreme and incomplete views of the face. An example of this is shown in the bottom left of Figure~\ref{fig:teaser}. This problem is further complicated in cases where the sensors measure a modality that differs from that used to build the model. A common example is the use of IR cameras, which defeats naive matching with images captured in the visible spectrum due to pronounced sub-surface scattering in IR. To address this problem, we leverage the property of CVAE, where weight sharing across similar modalities tends to preserve semantics. Specifically, we found that learning a common CVAE  over headset images and re-rendered versions of the multiview capture images allows us to correspond the two modalities through their common latent representation. This, in turn, implies correspondence from headset images to latent codes of the deep appearance model, over which we can learn a regression. Coupled with the deep appearance model, this approach allows the creation of a completely personalized end-to-end system where a person's facial state is encoded with the tracking VAE encoder and decoded with the rendering VAE decoder without the need for manually defined correspondences between the two domains. 

This paper makes two major contributions to the problem of real-time facial rendering and animation:
\begin{itemize}
\item A formulation of deep appearance models that can generate highly realistic renderings of the human face. The approach does not rely on high-precision rendering elements, can be built completely automatically without requiring any manual cleanup, and is well suited to real-time interactive settings such as VR. 
\item A semi-supervised approach for relating the deep appearance model latent codes to images captured using sensors with drastically differing viewpoint and modality. The approach does not rely on any manually defined correspondence between the two modalities but can learn a direct mapping from images to latent codes that can be evaluated in real-time. 
\end{itemize}
In \S\ref{sec:RelatedWork} we cover related work, and describe our capture apparatus and data preparation in \S\ref{sec:CapturingFacialData}. The method for building the deep appearance models is described in \S\ref{sec:model}, and the technique for driving it from headset mounted sensors in \S\ref{sec:DrivingDataDrivenAvatar}. Qualitative and quantitative results from investigating various design choices in our approach are presented in \S\ref{sec:Results}. We conclude in \S\ref{sec:Discussion} with a discussion and directions of future work.  

%This paper describes a data-driven method for rendering photorealistic human faces in virtual reality enabled by the following contributions:
%\begin{itemize}
%\item A novel pathway for creating pixels by learning how appearance varies according to view direction by considering \emph{view-specific} appearance maps;
%\item A view of facial rendering as the problem of predicting \emph{view-dependent} appearance maps;
%\item Replacing the traditional shading process with a learned function that predicts \emph{view-dependent} appearance maps of the face;
%\item A formulation of deep active appearance models that jointly models the variation of geometry and texture and enables reproduction of realistic facial appearance with deep networks while also retaining the benefits of the traditional graphics pipeline;
%\item A method to drive these models from headset-mounted cameras that relies on generating correspondence between headset data and multi-camera capture data completely unsupervised.
%\end{itemize}

\begin{figure*}[t]
    \begin{center}
    \includegraphics[width=1.0\textwidth]{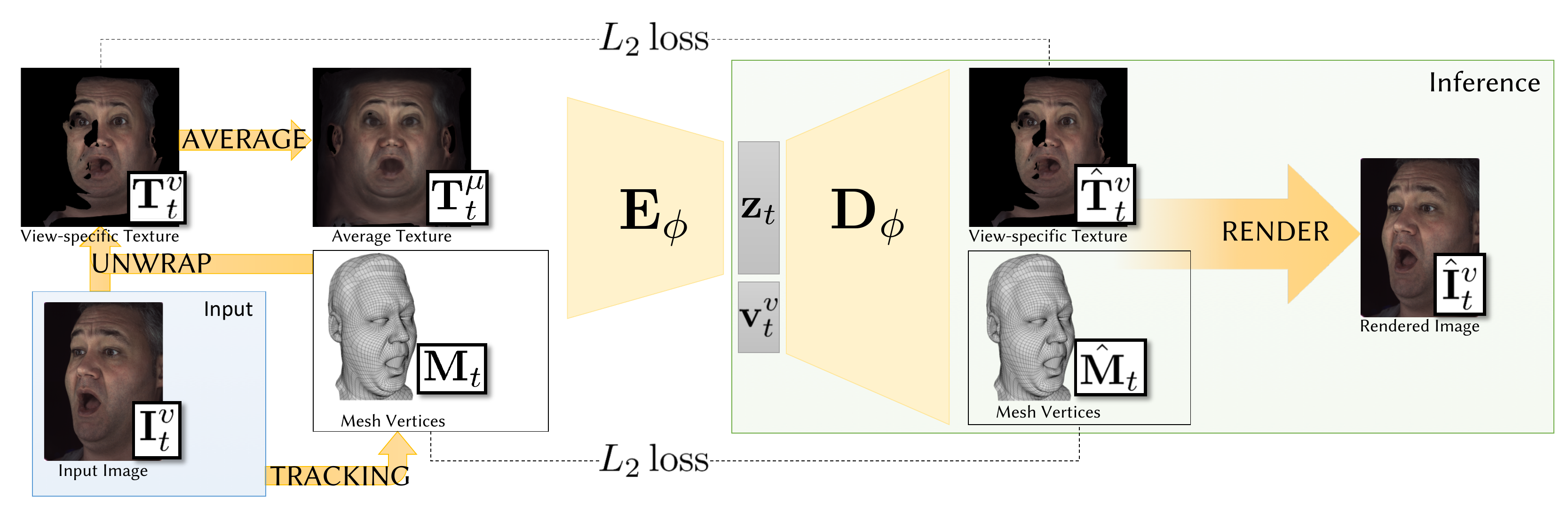}
    \end{center}
\caption{Pipeline of our method. Our method begins with input images from the multi-camera capture setup. Given a tracked mesh, we can unwrap these images into view-specific texture maps. We average these texture maps over all cameras for each frame and input it and the mesh to a variational autoencoder. The autoencoder learns to reconstruct a mesh and view-specific texture because the network is conditioned on the output viewpoint. At inference time, we learn to encode a separate signal into the latent parameters $\mathbf{z}$ which can be decoded into mesh and texture and rendered to the screen.}
\label{fig:pipeline}
\end{figure*}

\section{Related Work}
\label{sec:RelatedWork}

% more things to cite potentially:
% Shu et al. estimate intrinsic images (albedo map, normal map, illumination) from a single image using deep networks \cite{Shu:CVPR:2017}.
% Face2face?
% traditional parameterizations?
% * 'Meet Mike' ?

% GAN
% original gan, wgan, began, progressive growing gan

%* Human face inverse rendering work *
%\textbf{Inverse Rendering}.
%Many authors rely on some physical-inspired model of light, reflectance, and geometry to infer these latent variables from images \cite{Aldrian:TPAMI:2013,Kim:2017,Richardson:CVPR:2017}. These simplified assumptions often work well for skin but fail for areas like the eyes and inner mouth.

% * Active Appearance Model history
Active Appearance Models (AAMs)~\cite{Cootes:TPAMI:2001,Edwards:ICFGR:1998,Matthews:IJCV:2004,Tzimiropoulos:ACCV:2013} and 3D morphable models (3DMM)~\cite{Blanz:1999:MMS:311535.311556, Knothe2011} have been used as an effective way to register faces in images through analysis-by-synthesis using a low dimensional linear subspace of both shape and appearance. Our deep appearance model can be a seen as a re-purposing of the generator component of these statistical models for rendering highly realistic faces. To perform this task effectively, there are two major modifications that were necessary over the classical AAM and 3DMM approaches; the use of deep neural networks in-place of linear generative functions, and a system for view-conditioning to specialize the generated image to the viewpoint of the camera. 

%as opposed to registration, for which they were initially designed. 

%Although AAMs have fallen out of favor in recent years, we believe that the statistical separation and joint modeling of shape and appearance is a crucial element for creating highly realistic facial models that we can extend with deep networks.

%The main idea of these models is to jointly model the variation of geometry and texture of the face with Principal Component Analysis (PCA) to register facial pose and expression from images. 

% * Deep networks/modern methods
Recently, there has been a great deal of work using deep networks to model human faces in images. Radford et al.\ \citeyear{Radford:CoRR:2015} use Generative Adversarial Networks (GANs) to learn representations of the face and learn how movements in this latent space can change attributes of the output image (e.g., wearing eyeglasses or not). Hou et al.\ \citeyear{Hou:WACV:2017} use a variational autoencoder to perform a similar task. Kulkarni et al.\ \citeyear{Kulkarni:NIPS:2015} attempt to separate semantically meaningful characteristics (e.g., pose, lighting, shape) from a database of rendered faces semi-automatically. All these methods operate only in the image domain and are restricted to small image sizes (256$\times$256 or less). By contrast, we learn to generate view-specific texture maps of how the face changes due to facial expression and viewing direction. This greatly simplifies the learning problem, achieving high-resolution realistic output as a result.

% papers for generating avatars from images/image-based modeling of avatars
There has also been work on automatically creating image-based avatars using a geometric facial model \cite{Ichim:ToG:2015,Hu:ToG:2017,Thies:ToG:2018}. Cao et al.~\citeyear{Cao:SIGGRAPH:2016} propose an image-based avatar from 32 images captured from a webcam. To render the face, they construct a specially crafted function that blends the set of captured images based on the current facial expression and surface normals, which captures the dependence of both expression and viewpoint on facial appearance. Casas et al.~\citeyear{Casas:CASA:2016} propose a method for rapidly creating blendshapes from RGB-D data, where the face is later rendered by taking a linear combination of the input texture maps. Although both methods take an image-based approach, our method learns how appearance changes due to expression and view rather than prescribing it. Most importantly, these approaches are geared towards data-limited scenarios and, as a result, do not scale well in cases where data is more abundant. Our approach is designed specifically to leverage large quantities of high quality data to achieve the compelling depiction of human faces.

%* View-dependent texture mapping / spatially-varying BRDF / image-based rendering *
Because our method is concerned with view-dependent facial appearance, our work has a connection to others that have studied the view- and light-dependent appearance of materials and objects. One way to view this work is as a compression of the ``8D reflectance function that encapsulates the complex light interactions inside an imaginary manifold via the boundary conditions on the surface of the manifold'' \cite{Klehm:EUROGRAPHICS:2015}. That is, we are capturing the lightfield at a manifold on which the cameras lie and compressing the appearance via a deconstruction of texture and geometry. Image-based rendering methods \cite{Kang:ICIP:2000,Gortler:SIGGRAPH:1996} represent the light field non-parametrically and attempt to use a large number of samples to reconstruct novel views. The Bidirectional Texture Function \cite{Dana:ToG:1999} is a method to model both spatially-varying and view-varying appearance. These approaches have served as inspiration for this work.

To drive our model we develop a technique for learning a direct end-to-end mapping from images to deep appearance model codes. Although state of the art techniques for face registration also employ regression methods~\cite{Xiong-2013-7701,kazemi2014one}, most approaches assume a complete view of the face where correspondences are easily defined through facial landmarks. An approach closest to our work is that of Olszewski et al.~\citeyear{olszewski2016high}, which employ direct regression from images to blendshape coefficients. The main difference to our work is in how corresponding expressions are found between the two domains. Whereas in Olszewski et al.'s work, correspondences are defined semantically through peak expressions and auditory aligned features of speech sequences, our approach does not require any human defined correspondence, or even that the data contains the same sequence of facial motions. Since we leverage unsupervised domain adaptation techniques, the only requirement of our method is that the distribution of expressions in both domains are comparable. 

% domain adaptation papers
There have recently been a number of works that perform unsupervised domain adaptation using deep networks \cite{Liu:NIPS:2017,Kim:ICML:2017,Bousmalis:CVPR:2017,Zhu:ICCV:2017}. Many of these methods rely on Generative Adversarial Networks (GANs) to translate between two or more domains, sometimes with additional losses to ensure semantic consistency. Some works also use VAE in conjunction with other components to learn a common latent space between domains~\cite{Liu:NIPS:2017,Hsu:NIPS:2017}. Our approach differs from these works in that we leverage the pre-trained rendering VAE to help constrain the structure of the common latent space, resulting in semantics that are well matched between the modalities. 

%We take a simplified approach and use a conditional VAE to learn a common representation between domains by exploiting the disentangling behavior of VAE \cite{Higgins:ICLR:2017}.

\section{Capturing Facial Data}
\label{sec:CapturingFacialData}

% Refer capture setup
% Refer to Hao Li Nvidia paper that does similar thing
% Say in-house method
% Maybe ask Chenglei or Takaaki or Tomas to write up this section

% High-level
To learn to render faces, we need to collect a large amount of facial data. For this purpose, we have constructed a large multi-camera capture apparatus with synchronized high-resolution video focused on the face. The device contains 40 machine vision cameras capable of synchronously capturing 5120$\times$3840 images at 30 frames per second. All cameras lie on the frontal hemisphere of the face and are placed at a distance of about one meter from it. We use zoomed in 50mm lenses, capturing pore-level detail, where each pixel observes about 50$\mu$m on the face.

% Lighting
We preprocess the raw video data by performing multiview stereo reconstruction. In order to achieve the best results, we evenly place 200 directional LED point lights directed at the face to promote uniform illumination. 

% Capture script
To keep the distribution of facial expressions consistent across identities, we have each subject make a predefined set of 122 facial expressions. Each subject also recites a set of 50 phonetically-balanced sentences. The meta-data regarding the semantic content of the recordings is not utilized in this work, but its inclusion as additional constraints in our system is straightforward and is a potential direction for future work. 

% Tracking
As input to our deep appearance model learning framework, we take the original captured images as well as a tracked facial mesh. To generate this data, we build personalized blendshape models of the face from the captured expression performances, similar to~\citet{Laine:SCA:2017}, and use it to track the face through the captured speech performances by matching the images and the dense 3D reconstructions.

% 3D reconstruction,
%To densely reconstruct the 3D surface of the face, we first use the fast PatchMatch algorithm \cite{Galliani:ICCV:2015} to compute depth maps for each camera view. We then use Poisson surface reconstruction to merge the depth maps into a meshed surface.

% blendshape tracking,
%Unfortunately, the dense 3D reconstructions have no temporal consistency. However, we can use the 3D reconstructions and the image appearance to robustly track the face through time. To do this, we first form a blendshape basis by manually fitting a template mesh to the dense reconstructions of each facial expression image.
% TODO: There are some other steps that I don't remember
%We then perform blendshape tracking using this personalized blendshape basis by fitting to both the 3D reconstruction surface as well as the image appearance.
% TODO: There are some other details that I don't remember

\section{Building A Data-driven Avatar}\label{sec:model}

% Desired model properties
Our primary goal is to create extremely high-fidelity facial models that can be built automatically from a multi-camera capture setup and rendered and driven in real time in VR (90Hz). In achieving this goal, we avoid using hand-crafted models or priors, and instead rely on the rich data we acquired from our multiview capture apparatus.

% High-level method
We unify the concepts of 3D morphable models, image-based rendering, and variational autoencoders to create a real-time facial rendering model that can be learned from a multi-camera capture rig. The idea is to construct a variational autoencoder that jointly encodes geometry and appearance. In our model, the decoder outputs view-dependent textures---that is, a texture map that is ``unwrapped'' from a single camera image. It is view-specific and therefore contains view-dependent effects such as specularities, distortions due to imperfect geometry estimates, and missing data in areas of the face that are occluded.

% Conditioning on view
The critical piece of the proposed method is that we use a conditional variational autoencoder to condition on the viewpoint of the viewer (at training time, the viewer is the camera from which the texture was unwrapped; at test time, the viewpoint we want to render from; in both cases the direction is composed with the inverse of the global head-pose in the scene), allowing us to output the correct view-specific texture. At test time, we can execute the decoder network in real-time to regress from latent encoding to geometry and texture and finally render using rasterization.

\begin{figure*}[t]
    \begin{center}
        \includegraphics[width=1.0\textwidth]{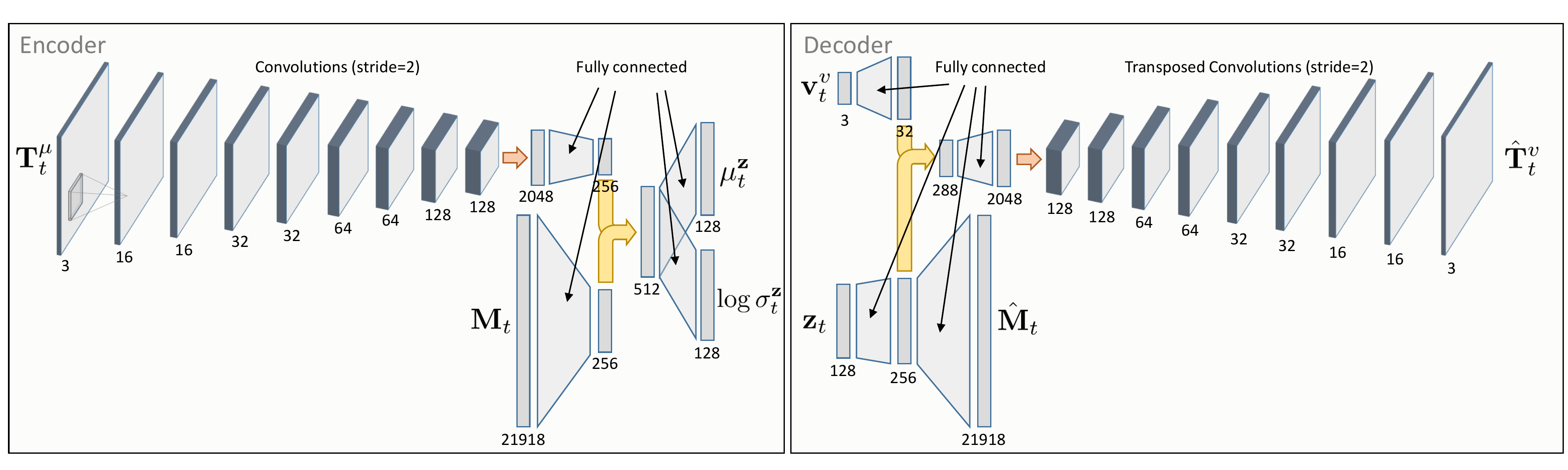}
    \end{center}
    \caption{Architecture of the Encoder and Decoder. The textures $\mathbf{T}_t^{\mu}$ and $\mathbf{T}_t^{v}$ are 3-channel 1024 $\times$ 1024 images. Each convolutional layer has stride $2$ and the number of channels doubles after the first convolution every two layers. We combine the texture and mesh subencodings via concatenation. The decoder runs these steps in reverse: we split the network into two branches and use transposed convolutions of stride $2$ to double the image resolution at every step. This decoder network executes in less than 5 milliseconds on an NVIDIA GeForce GTX 1080 GPU.}
    \label{fig:decoder_arch}
\end{figure*}

% Pipeline overall
Figure \ref{fig:pipeline} visualizes the training and inference pipeline of our algorithm. For this work, we assume that coarse facial geometry has been tracked and we use it as input to our algorithm with the original camera images. After geometry is tracked, we ``unwrap'' texture maps for each camera and for every frame of capture. Unwrapping is performed by tracing a ray from the camera to each texel of the texture map and copying the image pixel value into the texel if the ray is not occluded. These view-specific texture maps are what we want to generate at test time: reproducing them will cause the rendered image to match the ground truth image after rendering. To learn how to generate them efficiently, we use a conditional variational autoencoder (CVAE) \cite{Kingma:CoRR:2013}. Because we jointly model geometry and appearance, our autoencoder has two branches: an RGB texture map and a vector of mesh vertex positions.

% Formal definition of method: inputs
Let $\mathbf{I}^{v}_{t}$ be an image from the multi-camera capture rig at time instant $t$ from camera $v$ (for a total of $\mathcal{V}=40$ cameras). In this work, we assume that the viewpoint vector is relative to the rigid head orientation that is estimated from the tracking algorithm. At each time instant we also have a 3D mesh $\mathbf{M}_t$ (7306 vertices $\times\, 3=21918$-dimensional vector)  with a consistent topology across time. Using the image and mesh we can ``unwrap'' a view-specific texture map $\mathbf{T}^{v}_{t}$ by casting rays through each pixel of the geometry and assigning the intersected texture coordinate to the color of the image pixel. During training, the conditional VAE takes the tuple $(\mathbf{T}^{\mu}_{t}, \mathbf{M}_t)$ as input and $(\mathbf{T}^{v}_{t}, \mathbf{M}_t)$ as the target. Here,
\begin{equation}
\mathbf{T}^{\mu}_{t} = \frac{\sum^{\mathcal{V}}_{v=1} w^{v}_{t} \odot \mathbf{T}^{v}_{t}}{\sum^{\mathcal{V}}_{v=1} w^{v}_{t}},
\end{equation}
is the average texture, where $w^{v}_t$ is a factor indicating whether each texel is occluded (0) or unoccluded (1) from camera $v$ and $\odot$ represents an element-wise product. The primary reason for this asymmetry is to prevent the latent space from containing view information and to enforce a canonical latent state over all viewpoints for each time instant. The effects of this is discussed below.

% Formal definition: VAE
The cornerstone of our method is the conditional variational autoencoder that learns to jointly compress and reconstruct the texture $\mathbf{T}^{v}_t$ and mesh vertices $\mathbf{M}_t$. This autoencoder can be viewed as a generalization of Principal Component Analysis (PCA) in conventional AAMs. The autoencoder consists of two halves: the encoder $\mathbf{E}_{\phi}$ and the decoder $\mathbf{D}_{\phi}$. The encoder takes as input the mesh vertices and texture intensities and outputs parameters of a Gaussian distribution of the latent space,
\begin{equation}
\mathbf{\mu}^{\mathbf{z}}_t, \log \mathbf{\sigma}^{\mathbf{z}}_t \leftarrow \mathbf{E}_{\phi}(\mathbf{T}^{\mu}_t, \mathbf{M}_t),
\end{equation}
where the function $\mathbf{E}_{\phi}$ is parameterized using a deep neural network with parameters $\phi$. At training time, we sample from the distribution,
\begin{equation}
\mathbf{z}_t \sim \mathcal{N}(\mathbf{\mu}^{\mathbf{z}}_t, \mathbf{\sigma}^{\mathbf{z}}_t),
\end{equation}
and pass it into the decoder $\mathbf{D}_{\phi}$ and compute the reconstruction loss. This process approximates the expectation over the distribution defined by the encoder. The vector $\mathbf{z}_t$ is a data-driven low-dimensional representation of the subject's facial state. It encodes all aspects of the face, from eye gaze direction, to mouth pose, to tongue expression.

% Formal definition: decoder
The decoder takes two inputs: the latent facial code $\mathbf{z}_t$ and the view vector, represented as the vector pointing from the center of the head to the camera $\mathbf{v}_t$ (relative to the head orientation that is estimated from the tracking algorithm). The decoder transforms the latent code and view vector into reconstructed mesh vertices and texture,
\begin{equation}
\hat{\mathbf{T}}^{v}_t, \hat{\mathbf{M}}_t \leftarrow \mathbf{D}_{\phi}(\mathbf{z}_t, \mathbf{v}^{v}_t),
\end{equation}
where $\hat{\mathbf{T}}^{v}_t$ is the reconstructed texture and $\hat{\mathbf{M}}_t$ is the reconstructed mesh. After decoding, the texture, mesh, and camera pose information can be used to render the final reconstructed image $\hat{\mathbf{I}}^{v}_t$.

% Architecture figure
Figure \ref{fig:decoder_arch} shows the architecture of the encoder and decoder. Conditioning is performed by concatenating the conditioning variable to the latent code $\mathbf{z}$ after each are transformed by a single fully connected layer. Since the mesh should be independent of viewpoint, it is only a function of the latent code. The texture decoder subnetwork consists of a series of strided transposed convolutions to increase the output resolution. In each layer, reparameterize the weight tensor using Weight Normalization \cite{Salimans:NIPS:2016}. We use leaky ReLU activations between layers with 0.2 leakiness. The decoder network must execute in less than 11.1 milliseconds to achieve 90Hz rendering for real-time VR systems. We are able to achieve this using transposed strided convolutions even with a final texture size of 1024 $\times$ 1024. This is a major departure from most previous work for generating facial imagery that has been limited to significantly smaller output sizes.

% Architecture details / Exploiting non-stationary statistics
Texture maps have non-stationary statistics that we can exploit in our network design. For example, DeepFace \cite{Taigman:CVPR:2014} utilizes ``locally-connected layers'', a generalization of convolution where each pixel location has a unique convolution kernel. We utilize a similar but simpler approach: each convolutional layer has a bias that varies with both channel \emph{and} the spatial dimensions. We find that this greatly improves reconstruction error and visual fidelity.

% Training / loss
To train our system, we minimize the L$_2$-distance between the input texture and geometry and the reconstructed texture and geometry plus the KL-divergence between the prior distribution (an isotropic Gaussian) and the distribution of the latent space:
\begin{align}
\begin{split}
\ell(\phi) = \sum_{v,t} & \lambda_T \left\lVert w^{v}_t \odot \left(\mathbf{T}^{v}_t - \hat{\mathbf{T}}^{v}_t\right)\right\rVert^2 + \lambda_M \Big\lVert\mathbf{M}_t - \hat{\mathbf{M}}_t\Big\rVert^2 + \\
& \lambda_Z \ \mathrm{KL}\Big(\mathcal{N}\left(\mathbf{\mu}^{\mathbf{z}}_{t}, \mathbf{\sigma}^{\mathbf{z}}_{t}\right) \,\big\|\, \mathcal{N}\big(0, \mathbf{I}\big)\Big),
\end{split}
\end{align}
where $w^{v}_t$ is a weighting term to ensure the loss does not penalize missing data (i.e., areas of the face that are not seen by the camera) and $\lambda_{*}$ are weights for each term (in all experiments we set these values to $\lambda_T{=}1.0$, $\lambda_M{=}1.0$, $\lambda_Z{=}0.01$). We use the Adam algorithm \cite{Kingma:CoRR:2013} to optimize this loss. Before training, we standardize the texture and geometry so that they have zero mean and unit variance. For each subject, our dataset consists of around 5,000 frames of capture under 40 cameras per frame. Training typically takes around one day for 200,000 training iterations with a mini-batch size of 16 on an NVIDIA Tesla M40.

% Inference / stereo textures
During test time, we execute the decoder half to transform facial encodings $\mathbf{z}$ and viewpoint $\mathbf{v}$ into geometry and texture. Using the architecture shown in Figure \ref{fig:decoder_arch}, this takes approximately 5 milliseconds on an NVIDIA GeForce GTX 1080 graphics card; well within our target of 11.1 milliseconds. In practice, we find that in VR it is desirable to decode twice, creating one texture for each eye to induce parallax also in the generated texture. Our network is able to generalize in viewpoint enough that the small difference in viewpoint between the two eyes improves the experience by giving an impression of depth in regions that are poorly approximated by the sparse geometry (e.g., the oral cavity).

% Average texture input vs view-specific texture input
For our viewpoint conditioning to work, the decoder network $\mathbf{D}_{\phi}$ must rely on the viewpoint conditioning vector to supply all the information about the viewpoint. In other words, the latent code $\mathbf{z}$ should contain \emph{no} information about the viewpoint of the input texture. This should be true for all types of variables that we may want to condition on at test time. The use of variational autoencoders \cite{Kingma:CoRR:2013} does promote these properties in the learned representation (as we will discuss in section \ref{sec:cond}). However, for the special case of viewpoint, we can explicitly enforce this factorization by supplying input $\mathbf{T}^{\mu}_t$ as the input texture, that is averaged across all cameras for a particular time instant $t$. This allows us to easily enforce a viewpoint-independent encoding of facial expression and gives us a canonical per-frame latent encoding.

\subsection{Conditioned Autoencoding}\label{sec:cond}

% Conditioning in general
A critical piece of our model is the ability to condition the output of the network on some property we want to control at test time. For our base model to work properly, we must condition on the viewpoint of the virtual camera. The idea of conditioning on some information we want to control at test time can be extended to include illumination, gaze direction, and even identity.  We can broadly divide conditioning variables into two categories: extrinsic variables (e.g., viewpoint, illumination, etc.) and intrinsic variables (e.g., speech, identity, gaze, etc.).

\subsubsection{Viewpoint Conditioning}

% View conditioning
The main variable that must be conditioned on is the virtual camera viewpoint. We condition on viewpoint so that at test time we will generate the appropriate texture from that viewer's point of view (relative to the avatars position and orientation). For this to work properly, the latent encoding $\mathbf{z}$ must not encode any view-specific information.

% viewpoint-independent expression
A viewpoint-independent expression representation can be achieved by using a variational autoencoder \cite{Kingma:CoRR:2013}, which encourages the latent encoding $\mathbf{z}$ to come from a zero-mean unit-variance Gaussian distribution. It has been observed that variational autoencoders tend to learn a minimal encoding of the data because of the regularization of the latent space \cite{Chen:CoRR:2016}.  Because of this, the network is encouraged to use the conditioning variable (in this case, viewpoint) as much as possible in order to minimize the number of dimensions used in the encoding. Rather than rely only on this mechanism, we input a texture map averaged over all viewpoints to the encoder network to ensure a viewpoint-independent latent encoding $\mathbf{z}$. We will, however, exploit this mechanism for other conditioning variables.

\subsubsection{Identity Conditioning}

% Motivation for identity conditioning
Avatars built with this system are person-specific. We would like to have a single encoder/decoder that models multiple people through an identity conditioning variable. With enough data, we may be able to learn to map single images to identity conditioning vectors, allowing us to dynamically create new avatars without needing a multi-camera capture rig.

% TODO: details on conditioning with 1 hot vector

% observations about independence of expression vector
Interestingly, with this identity conditioning scheme, we notice that the network learns an identity-independent representation of facial expression in $\mathbf{z}$ despite not imposing any correspondence of facial expression between different people. This is because the variational autoencoder (VAE) is encouraged to learn a common representation of all identities through its prior. Semantics tend to be preserved probably because the network can reuse convolutional features throughout the network.

%\subsection{Relighting}

% Simple trick for illumination
%Ideally, we would like our model to take the lighting conditions in addition to the viewer pose as input. Unfortunately, this requires an even more elaborate capture setup. To deal with this limitation, we propose a simple workaround: we consider the images captured by our setup to be albedo images (this is a very rough assumption). If we make this assumption, we can compute a reflectance map by simulating a Lambertian BRDF in the desired lighting environment. We find that this produces reasonable results. The main benefit of this approach is to make the object look like it fits into the environment rather than properly modeling complex lighting phenomena, like subsurface scattering.

%\begin{figure}
%    \begin{center}
%        Image goes here
%    \end{center}
%    \caption{Modeling low-frequency illumination. Here we simply consider the
%    output texture to be albedo and apply a reflectance map computed by
%    simulating a Lambertian surface in a set of different illumination
%    environments. The model retains its photorealistic apperance while also
%    looking appropriate in the environment.}
%    \label{fig:illum}
%\end{figure}

%Figure \ref{fig:illum} shows our results of applying this simple idea to several lighting environments. We can see that photorealism is retained while also making the avatar feel like it fits in with the environment.

\begin{figure*}[t]
    \begin{center}
        \includegraphics[trim={0 3mm 1mm 0},width=1.0\linewidth]{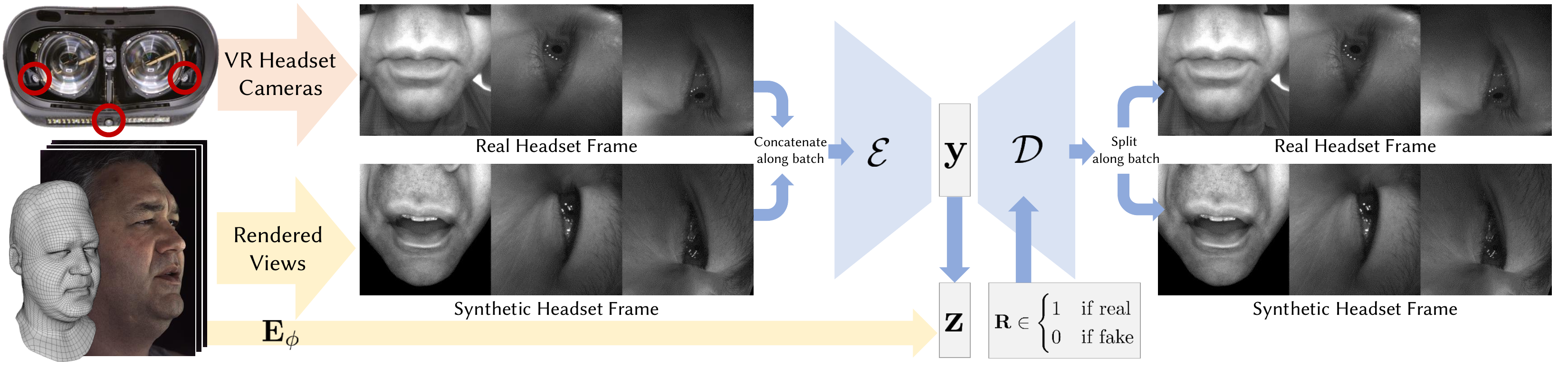}
    \end{center}
    \caption{Facial Tracking Pipeline. First, we generate synthetic headset images using image-based rendering on our multi-camera capture data. These images look geometrically like real headset images but not photometrically because of the difference in lighting. To account for this difference, we encode both synthetic headset images and real headset images using a VAE, which encourages learning a common representation of both sets. We can translate between the two modalities by flipping a conditioning variable.}
    \label{fig:track_pipeline}
\end{figure*}

\section{Driving a Data-Driven Avatar}
\label{sec:DrivingDataDrivenAvatar}

% TODO: need to introduce headset hardware somewhere

% Introduction to driving an avatar
Now that we are able to train the decoder to map latent encodings to texture and geometry, we can render animated faces in real-time. We would like the ability to drive our avatars with live or recorded performance rather than only playing back data from the original capture. In this section, we detail how to animate our avatar by controlling the latent facial code $\mathbf{z}$ with an unsupervised image-based tracking method. Because our focus is on creating the highest possible quality avatars and facial animation, we limit our scope to person-specific models (i.e., a person can drive only his or her own avatar).

% Getting correspondence
The primary challenge for driving performance with these avatars is obtaining correspondence between frames in our multi-camera capture system and frames in our headset. Note that this is a unique problem for virtual reality avatars because one cannot wear the headset during the multi-camera capture to allow us to capture both sets of data simultaneously, as the headset would occlude the face. We address this problem by utilizing unsupervised domain adaptation.

\subsection{Video-driven Animation}

% Image-based rendering for geometric transformation
Our approach centers around image-based rendering on the rich multi-camera data to simulate headset images. The first step is to compute approximate intrinsic and extrinsic headset camera parameters in the multi-camera coordinate system (we do this by hand for one frame and propagate the tracked head pose). Then, for each pixel of the synthetic headset image, we raycast into the tracked geometry and project that point into one of the multi-camera images to get a color value. This allows to produce synthetic images from the perspective of a headset with our multi-camera system.

% Crisis
Unfortunately, the lighting in the headset images and multi-camera images is quite different (and, in fact, of a different wavelength) so naively regressing from these synthetic headset images to facial encoding $\mathbf{z}$ will likely not generalize to real headset images.

% Related work: domain adaptation
There has been much work recently on performing unsupervised domain adaptation using adversarial networks that learn to translate images from one domain to another without any explicit correspondence (e.g., \cite{Zhu:ICCV:2017}). One possibility for us is to use an image-to-image translation approach to transform synthetic headset images into real headset images and then learn a regression from translated synthetic headset images to our latent code $\mathbf{z}$. This scheme has two main drawbacks: first, the network learning the regression to latent code $\mathbf{z}$ never trains on real headset images; second, the adversarial component of these methods tend to be difficult to train.

% Solution
We take the following alternative approach to solve this problem. First, we train a single variational autoencoder to encode/decode both real headset images and synthetic headset images. The Gaussian prior of the latent space will encourage the code $\mathbf{y}$ to form a common representation of both sets of images. We condition the decoder on a binary value indicating whether the image was from the set of real headset images or the set of synthetic images so that this information is not contained in the latent code. Next, we learn a linear transformation $\mathbf{A}_{\mathbf{y} \rightarrow \mathbf{z}}$ that maps the latent code $\mathbf{y}$ to the rendering code $\mathbf{z}$ for the synthetic headset images because we have correspondence between images of our multi-camera system $\mathbf{I}^{v}_{t}$ and latent codes $\mathbf{z}_{t} = \mathbf{E}_{\phi}(\mathbf{T}^{\mu}_{t}, \mathbf{M}_t)$ through our rendering encoder. If the VAE is successful in learning a common, semantically-consistent representation of real and synthetic headset images, then this linear regression will generalize to real headset images.

% semantic correspondence?
Note that while there is no guarantee that the semantics of the expression are the same when decoding in each of the two modalities, we observe that semantics tend to be preserved. We believe the primary reason for this is because the two image distributions are closely aligned and therefore the encoder network can make use of shared features.

% figure/details
Figure \ref{fig:track_pipeline} shows a pipeline of our technique. In this pipeline, the encoder $\mathcal{E}$ takes one headset frame $\mathbf{H}_t$ consisting of three images, mouth $\mathbf{H}^{m}_t$, left eye $\mathbf{H}^{l}_t$, and right eye $\mathbf{H}^{r}_t$. Each headset frame is either real $\mathbf{H}^{R}_t$ or synthetic $\mathbf{H}^{S}_t$. The encoder $\mathcal{E}$ produces a latent Gaussian distribution,
\begin{equation}
\mathbf{\mu}^{\mathbf{y}}_t, \log \mathbf{\sigma}^{\mathbf{y}}_t \leftarrow \mathcal{E}\left(\mathbf{H}_t\right).
\end{equation}
At training time, we sample from this distribution to get a latent code,
\begin{equation}
\mathbf{y}_t \sim \mathcal{N}\left(\mathbf{\mu}^{\mathbf{y}}_t, \mathbf{\sigma}^{\mathbf{y}}_t\right),
\end{equation}
as before. Our decoder $\mathcal{D}$ produces a headset frame given the latent code and an indicator variable :
\begin{equation}
\hat{\mathbf{H}}_t \leftarrow \mathcal{D}\left(\mathbf{y}_t, \mathbf{R}\right),
\end{equation}
where $\mathbf{R} \in \{0, 1\}$ indicates whether the decoder should decode a real headset frame or a synthetic headset frame. This indicator variable allows the latent code to contain no modality-specific information because the decoder network can get this information from the indicator variable instead. 

% architecture
The architecture of our headset encoder $\mathcal{E}$ is as follows: for each of the three types of headset images (lower mouth, left eye, right eye), we create a three-branch network with each branch containing eight stride-2 convolutions with each convolution followed by a leaky ReLU with 0.2 leakiness. The number of output channels for the first layer is 64 and doubles every other convolutional layer. The three branches are then concatenated together and two fully-connected layers are used to output $\mu^{\mathbf{y}}_{t}$ and $\log \sigma^{\mathbf{y}}_{t}$. For our experiments, the latent vector $\mathbf{y}$ has 256 dimensions. The decoder network $\mathcal{D}$ is similar: three branches are created, with each branch consisting of a fully-connected layer followed by eight stride-2 transposed convolutions with each layer followed by a leaky ReLU with 0.2 leakiness. The first transposed convolution has 512 channels as input and this value halves every other transposed convolution. We condition the decoder by concatenting the conditioning variable $\mathbf{R}$ to every layer, replicating across all spatial dimensions for convolutional layers. We don't use any normalization in these networks. Our encoder network runs in 10ms on an NVIDIA GeForce GTX 1080, which enables real-time tracking.

% Math pt. 2, training
% !! need to include lambdas for losses
% !! use subscripts to differentiate real/synthetic?
% !! introduce semantic loss?
% !! details of architecture
To train the network, we optimize the reconstruction loss, retargetting loss, and KL-divergence loss,
\begin{align}
\begin{split}
\ell(\theta) = \sum_t & \lambda_H \left\lVert \mathbf{H}_t - \hat{\mathbf{H}}_t \right\rVert^2 + \lambda_A \left\lVert \mathbf{z}_t - \mathbf{A}_{\mathbf{y} \rightarrow \mathbf{z}} \mathbf{y}_t \right\rVert^2 + \\
& \lambda_Y \mathrm{KL}\Big(\mathcal{N}\left(\mathbf{\mu}^{\mathbf{y}}_t, \mathbf{\sigma}^{\mathbf{y}}_t\right) \,\big\|\, \mathcal{N}\big(0, \mathbf{I}\big)\Big),
\end{split}
\end{align}
where $\mathbf{z}_t$ is known only for synthetic headset frames $\mathbf{H}^{S}$, $\mathbf{A}_{\mathbf{y} \rightarrow \mathbf{z}}$ linearly maps from tracking latent code $\mathbf{y}$ to rendering code $\mathbf{z}$, and $\lambda_{\cdot}$ are weights for each term of the loss (in all experiments we set $\lambda_H{=}1.0$, $\lambda_A{=}0.1$, and $\lambda_Y{=}0.1$). We optimize this loss using the Adam optimizer \cite{Kingma:ICLR:2014}, as before. To make our method robust to the position of the HMD on the face, we perform a random translation, scaling, and rotation crop on the images. We also randomly scale the intensity values of the images to provide some robustness to lighting variation.

% Conclusion
This completes the entire pipeline: a headset image is input to the headset encoding network $\mathcal{E}$ to produce the headset encoding $\mathbf{y}$, then it is translated to the multi-camera encoding $\mathbf{z}$, and finally it is decoded into avatar geometry $\hat{\mathbf{M}}$ and texture $\hat{\mathbf{T}}$ and rendered for the user. This architecture also works well for social interactions over a local- or wide-area network as only the latent code $\mathbf{z}$ needs to be sent across the network, reducing bandwidth requirements.

\begin{figure*}[t]
	\centering
    \includegraphics[width=1.0\textwidth]{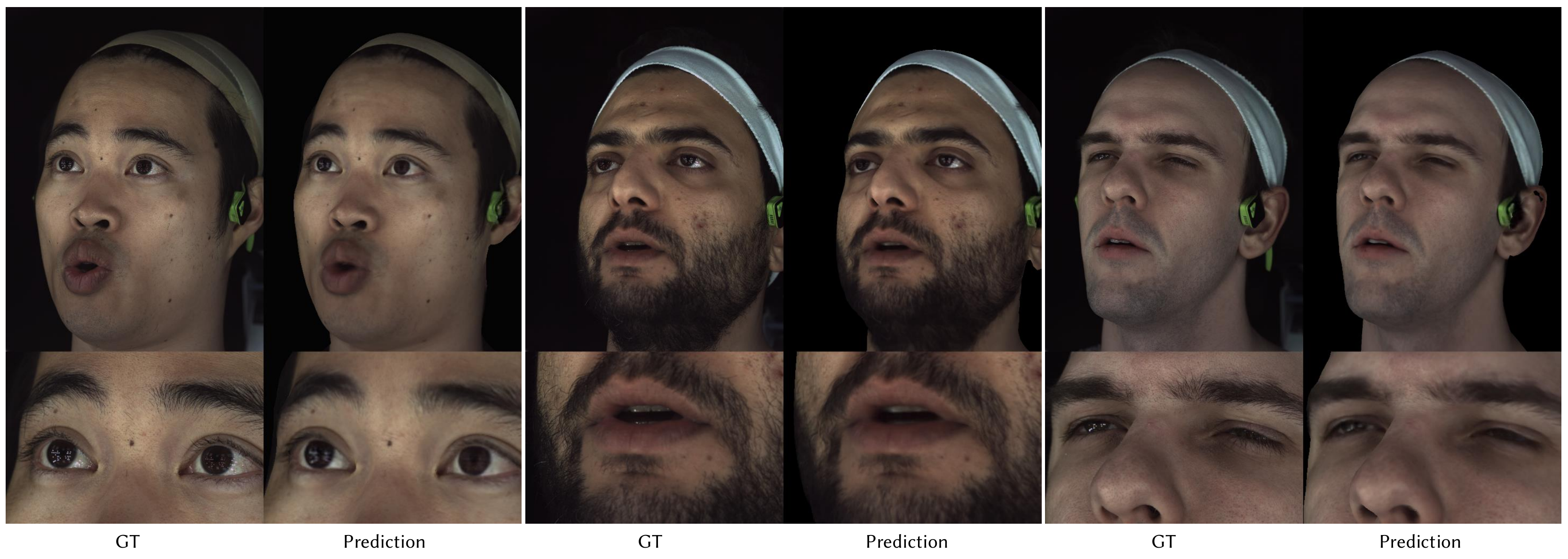}
    %\begin{center}
    %\begin{tabular}{ c c c }
    %    \includegraphics[width=0.32\textwidth]{figs/qual/002576793_rendergt_330033.png} &
    %    \includegraphics[width=0.32\textwidth]{figs/qual/002608483_rendergt_330033.png} &
    %    \includegraphics[width=0.32\textwidth]{figs/qual/002573267_rendergt_330033.png} \\
    %\end{tabular}
    %\end{center}
    \caption{Qualitative Results. This figure shows comparisons between the ground truth recordings and rendered mesh and texture predicted by our network (note that these frames were not in the training set) for four different subjects. We can see that our model achieves a high level of realism although it has some artifacts, in particular blurring within the mouth (we do not have tracked geometry inside the mouth, so our network must try to emulate that geometry with the texture). We find that these artifacts are less bothersome than traditional realtime human face renderings.}
    \label{fig:rendering_results}
\end{figure*}

% Video-driven Animation
%Video-based tracking can be achieved by performing blendshape-based tracking and then retargetting blendshapes to the latent code $\mathbf{z}$. The main idea is to run the tracker on the multi-camera capture data to learn a mapping from blendshapes to $\mathbf{z}$. At test time, we run the tracker on the driving capture device and map the tracking parameters to $\mathbf{z}$ using the mapping learned at training. After obtaining $\mathbf{z}$, we pass it through the decoder and render the resulting texture and geometry to the screen.

%\subsection{Audio-driven Animation}
%
%% Audio-driven Animation
%We also devise a method to animate our avatars with speech data. Our method has three parts: an audio backend, which computes phoneme probabilities from raw audio samples, a regression from phoneme probabilities to blendshape coefficients, and finally a linear retargetting layer from blendshape coefficients to latent code $\mathbf{z}$.
%
%% details
%% learn audio -> phoneme with temporal convolutional network
%% compute phonemes previous network, learn phoneme -> blendshape map
%% compute per-person blendshape -> z map
%
%% discussion

\begin{figure}[t]
    \centering
    \includegraphics[width=0.49\textwidth]{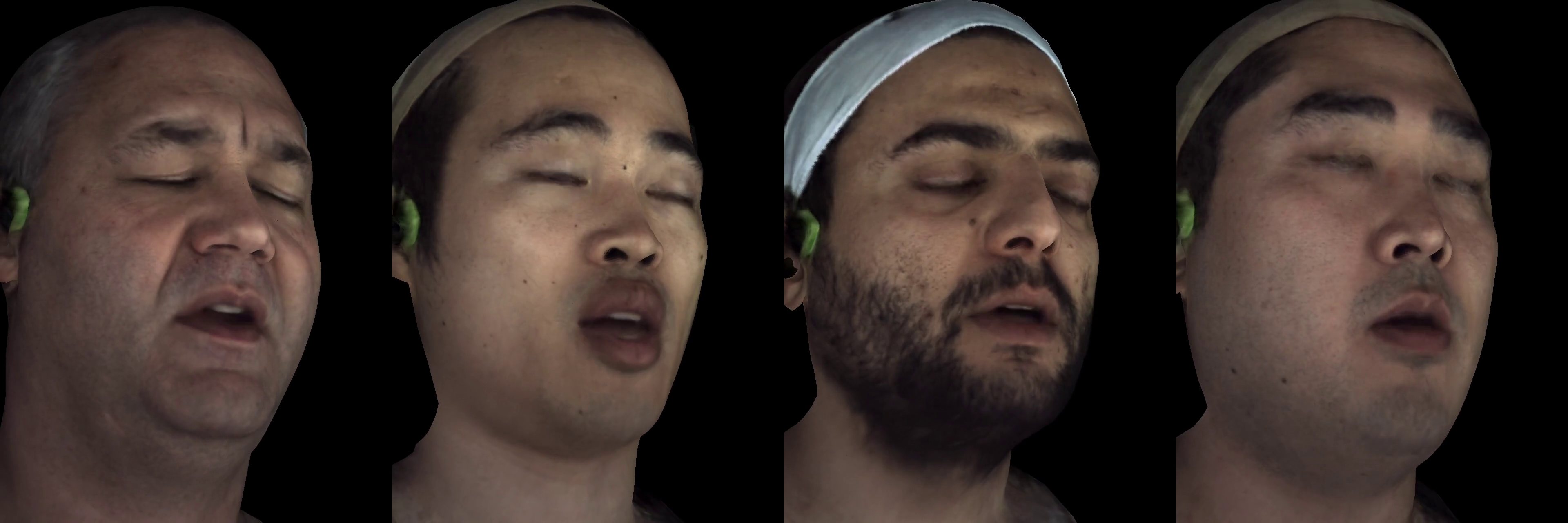}
    \caption{Identity Conditioning. Here we demonstrate the automatic expressions correspondence obtained through identity conditioning. Here we decode the same latent code $\mathbf{z}$ into different people by conditioning on separate identity one-hot vectors.}
    \label{fig:cond_ident}
\end{figure}

\begin{figure}[t]
    \centering
   	\includegraphics[width=0.49\textwidth]{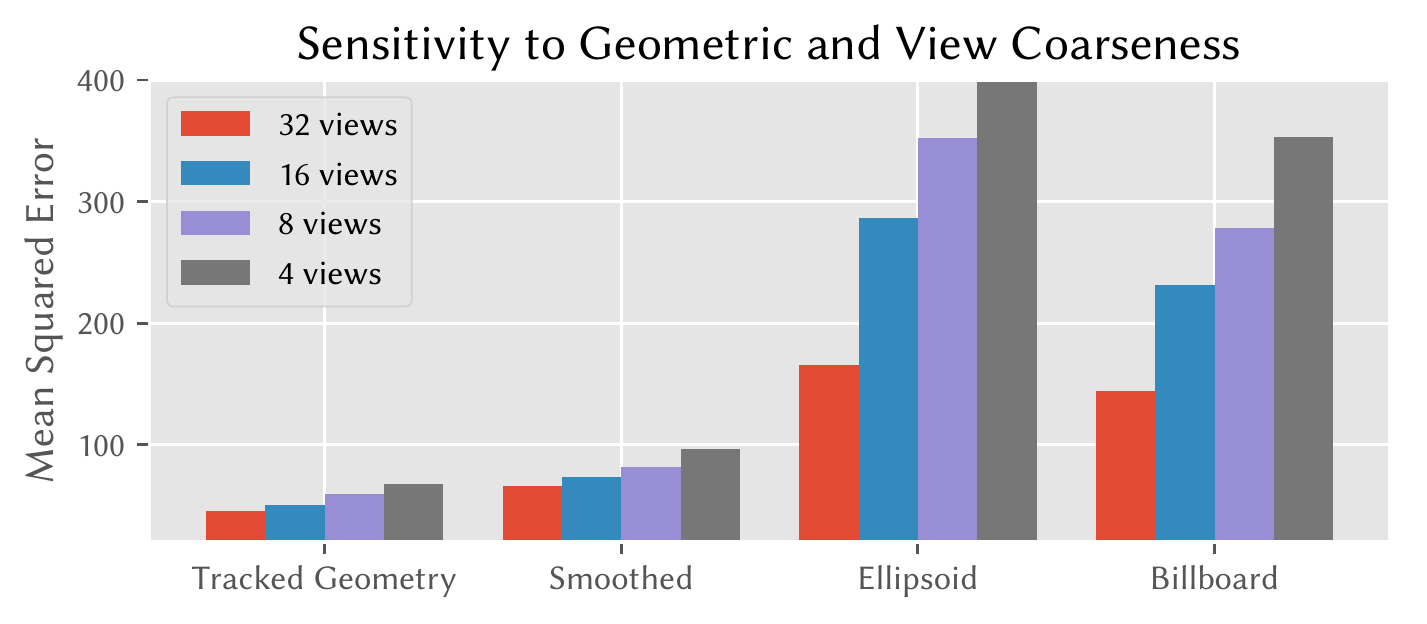}     
    \caption{Quantitative Effects of Geometric and Viewpoint Coarseness. We show the image-based mean-square error (MSE) for each geometry model and each set of training cameras (32, 16, 8, and 4 viewpoints). Here we see that accuracy of the geometric model allows the model to generalize well to the 8 validation viewpoints even when it has only seen 4 viewpoints at train time.}
    \label{fig:geomcoarseness}
\end{figure}

% !! use icon to represent different geometry types
\begin{figure*}[t]
    \centering
    \includegraphics[width=1.0\textwidth]{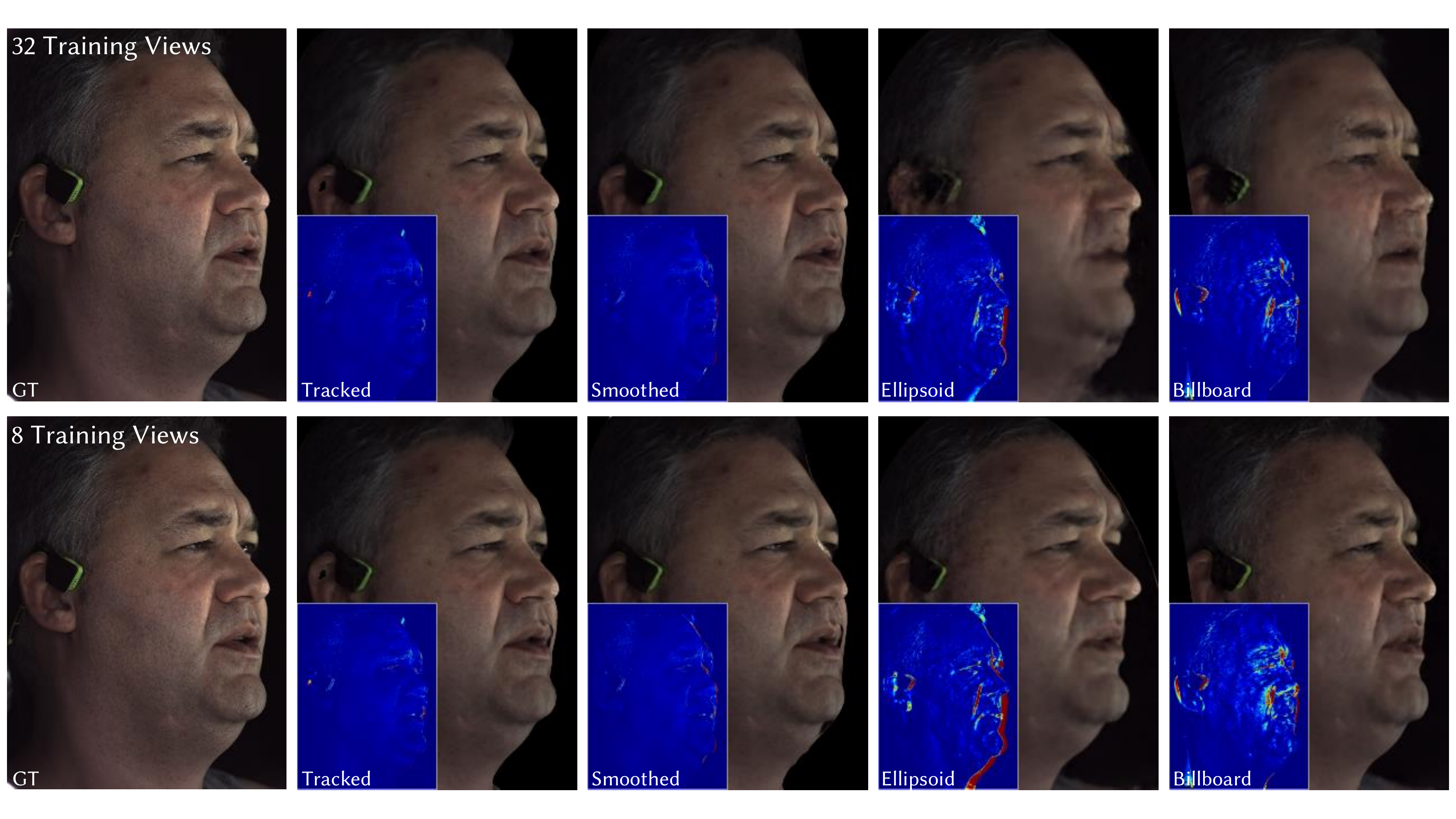}
    \caption{Qualitative Results for Evaluating the Effects of Geometric and Viewpoint Coarseness. We show renderings of the avatar trained with four different geometry models and with two different sets of viewpoints. Each render has an inset showing the squared difference image from the ground truth. Note that the viewpoint shown was not seen during training. We see that when geometry matches the true geometry well, the rendering model generalizes to new viewpoints (see row ``8 Training Views'', column ``Tracked''). When the geometry is too far from the true geometry, we start to see strange artifacts in the resulting image (see row ``32 Training Views'', column ``Billboard'').}
    \label{fig:geomcoarseness2}
\end{figure*}

\subsection{Artist Controls}

% Artist controls overview
In addition to driving the deep appearance model with video, we want to provide a system for rigging these models by character artists. To do this, we use a geometry-based formulation where we optimize to find a latent code that satisfies some geometric constraints, similar to~\citet{Lewis:CGA:2010}. The artist can click vertices and drag them to a new location to create a constraint.

% details
To find the latent code given the geometric constraints, we solve,
\begin{equation}
\arg\min_{\mathbf{z}} \sum^{N}_{i=1} \| \hat{\mathbf{M}}_i - \tilde{\mathbf{M}}_{i} \|^2 + \lambda_{P}\| \mathbf{z} \|^2,
\end{equation}
where $N$ is number of constraints,
$\hat{\mathbf{M}} = \mathbf{D}_{\phi}(\mathbf{z})$ is the mesh resulting from decoding $\mathbf{z}$, $\hat{\mathbf{M}}_i$ is the vertex of constraint $i$, and $\tilde{\mathbf{M}}_{i}$ is the desired location for the vertex of constraint $i$. The regularization on $\mathbf{z}$ can be seen as a Gaussian prior, which is appropriate because the variational autoencoder encourages $\mathbf{z}$ to take on a unit Gaussian distribution (we set $\lambda_P{=}0.01$). Even though the constraint is only placed on geometry, the model will produce a plausible appearance for the constraints because geometry and appearance are jointly modeled.

%\section{Results}

%\subsection{Effect of Geometric Coarseness}
%\begin{enumerate}
%\item Full Geometry
%\item Average 3D Face
%\item Ellipsoid
%\item Billboard/Sprites (Fronto-parallel Plane) [VAE-based; GAN-based]
%\end{enumerate}

%\subsection{Importance of Sampling Granularity of Viewpoints}
%\begin{enumerate}
%\item Number of Pixels vs Intensity levels (as the number of viewpoints decreases)
%\item plot distribution of error of uv map
%\item SNR (2D Theta vs Phi plots) as sampling increases
%\end{enumerate}

%\subsection{DAM vs AAM (how much does the nonlinearity help?)}
%\begin{enumerate}
%\item Linear
%\item Bilinear 
%\item Nonlinear
%\end{enumerate}

%Processing time, SNR

% Architecture experiments
%   * weight norm, batch norm
%   * how to condition

% Visualization of errors with more granularity
% 

% Evaluation of view-independent latent representation
%   * subset of experiments taking in view-specific texture

%\subsection{Image-based vs UV/Geometry-based Optimization Error}
%SNR (End to end, allows improvement of geometry inference)

%\subsection{Qualitative Results}
%\begin{itemize}
%\item High Resolution Output Image
%\item Sampling the latent space (How much of the latent is monster infested?)
%\item Audio-driven animation? [Audio: Iain]
%\item Keypoint-driven animation? [Tomas, Iain?]
%\item Registration? [Jason]
%\item Mugsy style collage
%\item Film strip of extreme expression
%\item Eye and Teeth "faking" results [Tomas]
%\end{itemize}

\section{Results}
\label{sec:Results}

In this section, we give quantitative and qualitative results of our method for rendering human faces and driving them from HMD cameras. First, we explore the effects of geometric coarseness, viewpoint sparsity, and architecture choice on avatar quality. Next, we demonstrate how our tracking variational autoencoder can preserve facial semantics between two modalities (synthetic and real) by switching the conditioning label. Finally, we show results of our complete pipeline: encoding HMD images and decoding and rendering realistic avatars.

\subsection{Qualitative Results}

% Rendering results
Figure \ref{fig:rendering_results} shows results of our method's renderings compared to ground truth images for three different subjects. Our model is able to capture subtle details of motion and complex reflectance behavior by directly modeling the appearance of the face. Although some artifacts are present (e.g., blurring inside the mouth), they are less distracting than other types of facial rendering artifacts.

Figure \ref{fig:cond_ident} shows results of our multi-identity model. For this experiment, we used data from 8 different people and trained one network conditioned on a latent identity vector. Currently this model cannot generate plausible interpolations of people, but we believe this will be possible with significantly more data.

% TODO: investigate more specific parts of the model qualitatively

\subsection{Effect of Geometric and Viewpoint Coarseness}

% Geometry vs. Texture / Texture compensation
A fundamental part of our model is that we jointly model geometry and texture. As a consequence, our model is able to ``correct'' for any bias in mesh tracking by altering the output of the texture maps to best match the true view-specific texture map. This powerful ability is part of what allows us to produce highly realistic avatars. There is a limit, however, to the ability of the texture network to resolve those errors. To explore this, we have constructed a series of experiments to analyze quantitatively and qualitatively how the model behaves when the geometry becomes coarser and viewpoints become sparser.

% Experimental details: full model, smoothed, ellipsoid, billboard
We investigate four different types of geometric coarseness: our original tracked geometry, significant smoothing of the original tracked geometry (we use 1000 iterations of Taubin smoothing for this \cite{Taubin:ICCV:1995}), a static ellipsoid subsuming the original mesh, and a billboard quad. We test billboard quad because image-based face generation in computer vision is typically performed on flat images. For each of these cases, we build our deep appearance model with different sets of training viewpoints (32 viewpoints, 16 viewpoints, 8 viewpoints, and 4 viewpoints) and compare performance in terms of mean-squared error on a set of 8 validation viewpoints held out from the total set of training viewpoints.

% Figure
% image-based reconstruction loss for each method, qualitative results
Figure \ref{fig:geomcoarseness} shows quantitative results for each of the four levels of geometric coarseness with four different sets of training viewpoints. The image-space mean-squared error (MSE), computed on the 8 validation viewpoints, shows a clear decrease in quality as the geometry is made more coarse. When the geometry input to our algorithm does not match the true geometry of the scene, the texture network must do more work to fill in the gaps, but there are limits to how well it generalizes. When the geometry fits well, the model performs well even with a small number of training viewpoints. When the geometry is too coarse, however, many viewpoints are needed for good generalization to new viewpoints.

Figure \ref{fig:geomcoarseness2} shows qualitative results for the same experiment. This figure shows that accurate geometry (column ``tracked'') generalizes even when there are few views for training. Coarse geometry, (columns ``ellipsoid'' and ``billboard''), however, generalizes poorly by introducing artifacts or by simply replicating the closest training viewpoint. Note that with coarse geometry the network has no problem learning the appearance of the training views but the generalization to new viewpoints is very poor.

\subsection{Architecture Search}

\begin{figure}[t]
    \begin{center}
        \includegraphics[width=0.49\textwidth]{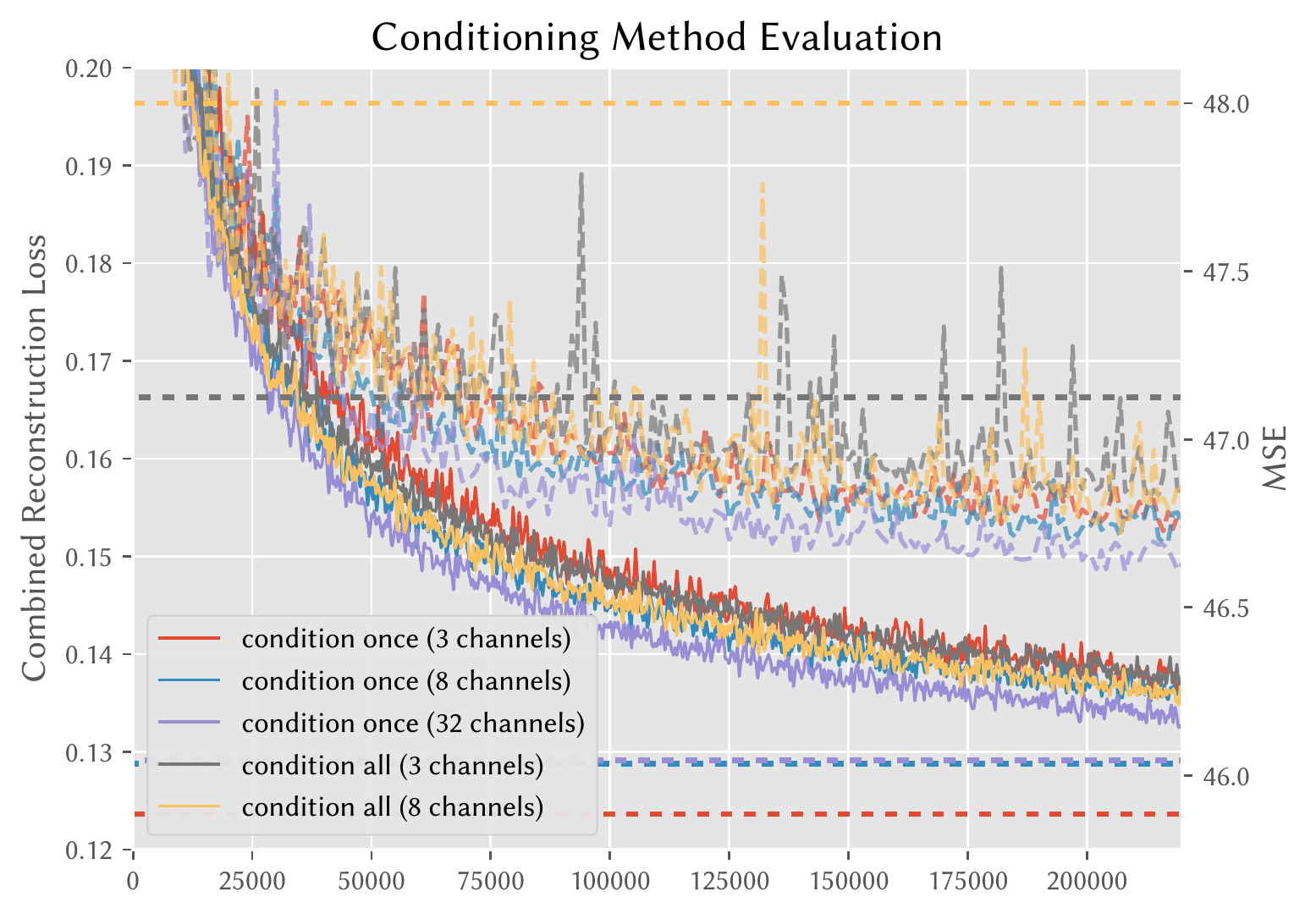}
    \end{center}
    \caption{Evaluation of Conditioning Method. We evaluate several ways to condition the network on viewpoint (red conditions the raw view vector on a single layer, blue conditions the view vector after a 3 $\times$ 8 fully-connected layer, purple conditions after a 3 $\times$ 32 fully-connected layer, grey conditions the raw view vector on all decoder layers, and yellow conditions the view vector after a 3 $\times$ 8 fully-connected layer on all decoder layers). Solid lines show training error (reconstruction loss), dashed lines show validation error (reconstruction loss), and flat dashed lines show image-space mean-squared error on a held-out set of cameras. Overall, conditioning on one early layer in the network tends to perform best.}
    \label{fig:condsearch}
\end{figure}

\begin{figure}[t]
    \begin{center}
        \includegraphics[width=0.49\textwidth]{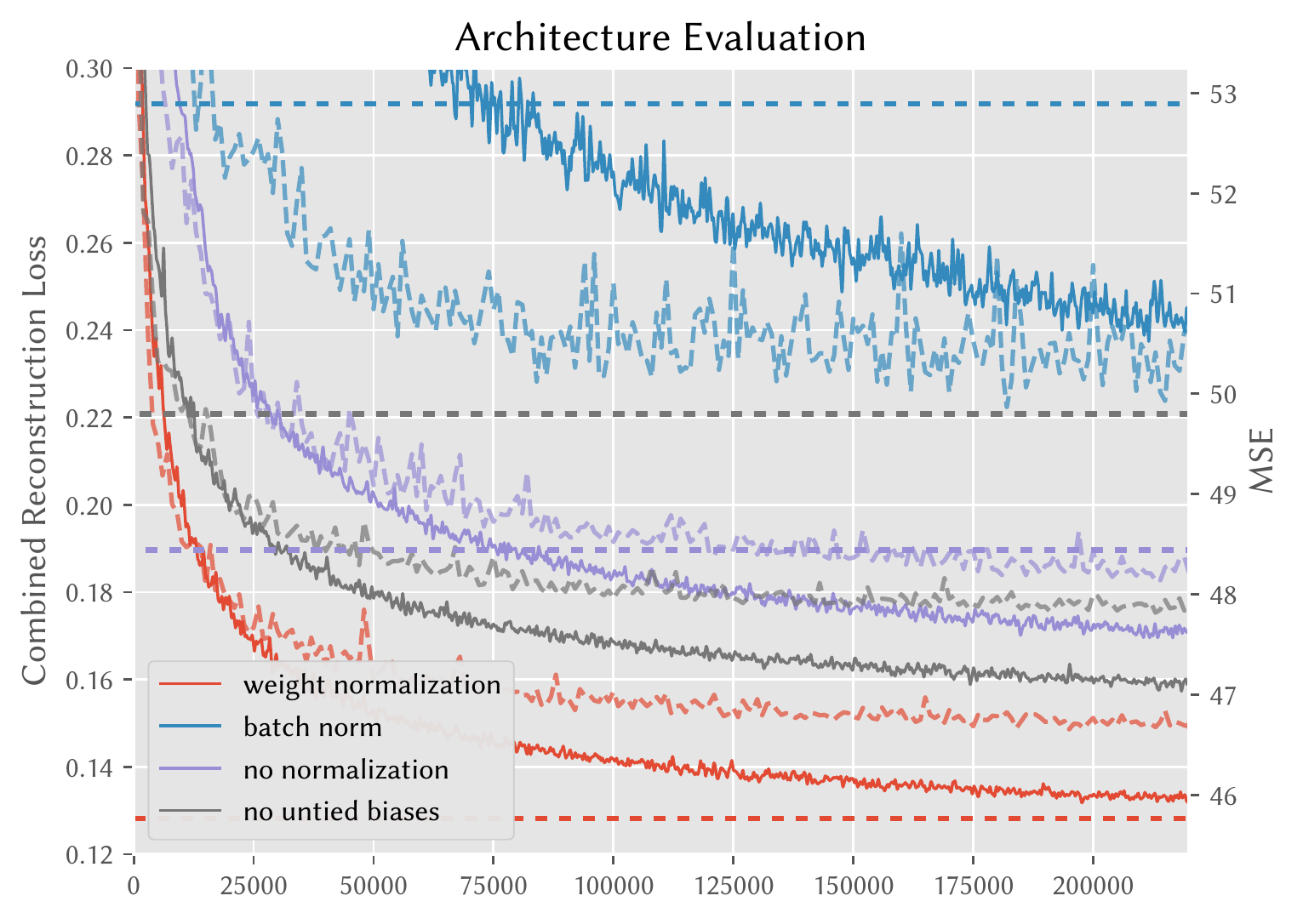}
    \end{center}
    \caption{Evaluation of Architectures. Solid lines show training error (reconstruction loss), dashed lines show validation error (reconstruction loss), and flat dashed lines show image-space mean-squared error on a held-out set of cameras. We found that using both weight normalization \cite{Salimans:NIPS:2016} and per-channel-and-pixel biases yields the highest quality avatar.}
    \label{fig:archsearch}
\end{figure}

% Intro/Experimental setup
We perform an evaluation of different architectures for our render encoder and decoder networks. We primarily tested across two axes: conditioning techniques and normalization methods. For conditioning techniques we tested two categories of conditioning: conditioning on an early decoder layer by concatenation and conditioning on all decoder layers by addition. For normalization techniques, we tested no normalization, batch normalization \cite{Ioffe:ICML:2015}, and weight normalization \cite{Salimans:NIPS:2016}. Finally, we investigate the advantage of a deep architecture over a linear or bilinear model. In each case, we report image-space mean-squared error for all valid head pixels. Here, head pixels are determined by raycasting into a dense stereo reconstruction of each frame.

\subsubsection{Conditioning Methods}

% Results
Figure \ref{fig:condsearch} shows the results of our conditioning evaluation. The figure shows the reconstruction loss and the MSE for a series of experiments changing the conditioning method. We evaluate across two main axes: expanding the view vector to a larger dimensionality with a fully connected layer before conditioning, and conditioning on one early layer versus conditioning on all layers. Here, we found that conditioning on a single early layer in the decoder overall seems to outperform conditioning all layers in terms of the image-based mean-squared error on the set of 8 validation views. We also found that transforming the view vector before conditioning seems to be beneficial.

\subsubsection{Normalization Methods}

% Results
Figure \ref{fig:archsearch} shows the results of our normalization evaluation. We tested three normalization methods: no normalization, batch normalization, and weight normalization. In addition, we tested weight normalization without using per-channel-and-pixel biases (described in section \ref{sec:model}). The figure shows that weight normalization far outperforms the other normalization techniques and that per-channel-and-pixel biases are very important for accuracy.

\subsubsection{Linear AAM vs.\ Deep AAM}

% Deep representations vs Linear
Aside from view conditioning, the main way we have augmented traditional AAMs is by changing the latent subspace from linear to nonlinear. The primary effect of this is to increase the expressiveness of the model although it also reduces the number of parameters. To evaluate the trade-off, we train our deep model compared to a linear model and a bilinear model (bilinear in view and facial state) with the same latent space dimensionality. Unfortunately a linear model requires $(21918 + 3 \cdot 1024^2) \cdot 128 \cdot 4 \cdot 2 = 3.24 \mathrm{GB}$ of weights so we use only a $512 \times 512$ output resolution for the linear and bilinear models.

%\begin{figure}[t]
%    \begin{center}
%        \includegraphics[width=0.48\textwidth]{figs/deepvslin.png}
%    \end{center}
%    \caption{Linear vs.\ Nonlinear representation. We evaluate our proposed method compared to a linear model (i.e., texture and geometry is a linear function of the latent code $\mathbf{z}$). Our model is not only better in terms of the image-based mean-squared error but also smaller in number of parameters (61MB vs. 411MB vs. 1.2GB).}
%    \label{fig:linear_nonlinear}
%\end{figure}
\begin{table}[t]
    \caption{Linear vs.\ Nonlinear representation. We evaluate our proposed method compared to a linear model (i.e., texture and geometry is a linear function of the latent code $\mathbf{z}$). Our model is not only better in terms of the image-based mean-squared error but also smaller in number of parameters (110MB vs. 806MB vs. 1.6GB).}
	\centering
	\begin{tabular}{ p{0.08\textwidth} p{0.10\textwidth} p{0.10\textwidth} p{0.10\textwidth} }
		\toprule
		& \multicolumn{3}{c}{Model Type} \\
		\cmidrule(l){2-4}
        & Our method & Linear & Bilinear \\
		\midrule
        Subject 1 & \textbf{39.12} & 59.00 & 69.17 \\
		\midrule
        Subject 2 & \textbf{87.67} & 121.44 & 128.94 \\
		\midrule
        Subject 3 & \textbf{128.07} & 173.33 & 168.16 \\
		\bottomrule
	\end{tabular}
\label{table:linear_nonlinear}
\end{table}

% Figure
Table \ref{table:linear_nonlinear} shows quantitative results for linear and nonlinear representations. For each subject and method, we give the image-space mean-squared error. Our model has clear advantages over linear and bilinear models, including compactness in terms of number of parameters and reconstruction performance.

\subsection{Image-based vs.\ Texture/Geometry-based Loss}
%SNR (End to end, allows improvement of geometry inference)

% Motivation for image-based loss
Traditional AAMs have been formulated as a PCA decomposition in terms of the texture and geometry jointly. By reformulating the AAM as an autoencoder we can actually make the loss a function of the reconstructed image $\hat{\mathbf{I}}^{v}_t$ itself. This is useful as it allows us to train on the metric on which we want to evaluate. It also opens the possibility of allowing the model to refine the output geometry over the input tracked geometry that may be erroneous.

% Formulation
To optimize with respect to the image-based reconstruction error, we rewrite the loss function as,
\begin{align}
\begin{split}
\ell(\phi) = \sum_{v,t} & \lambda_{I} \big\|m^{v}_t \odot \big(\mathbf{I}^{v}_t - \hat{\mathbf{I}}^{v}_t\big)\big\|^2 + \lambda_{T} \big\|w^{v}_t \odot \big(\mathbf{T}^{v}_t - \hat{\mathbf{T}}^{v}_t\big)\big\|^2 + \\ & 
\lambda_{M} \big\|\mathbf{M}_t - \hat{\mathbf{M}}_t\big\|^2 + \lambda_{Z} \mathrm{KL}\Big(\mathcal{N}(\mathbf{\mu}^{\mathbf{z}}_{t}, \mathbf{\sigma}^{\mathbf{z}}_{t}) \big\| \mathcal{N}(0, \mathbf{I})\Big),	
\end{split}
\end{align}
where $\hat{\mathbf{I}}^{v}_{t}$ is the reconstructed image from view $v$ at time-instant $t$, and $m^{v}_t$ is a mask that only penalizes pixels that contain the head (derived from the dense stereo reconstructions). %Using a differentiable rendering layer, we can backpropagate through the parameters of our model.
We create a semi-differentiable rendering layer using two steps: the first step (non-differentiably) rasterizes triangle indices to an image; the second step (differentiably) computes texel coordinates for each pixel given the triangle indices, camera parameters, and mesh vertices and samples the texture map at the coordinate computed for each pixel. This layer allows us to backpropagate through the parameters of our model and train the system end-to-end.

% Experimental setup
To evaluate how an image-based loss during training may improve the results we compare two different models: a model optimized with texture+geometry loss and a model optimized with texture+geometry+image loss. We found that without a geometry term, the loss becomes unstable and the geometry tends to lose its structure.

\begin{table}[t]
	\caption{Image MSE for Different Training Loss Combinations. We train with two different training objectives (TG--texture + geometry loss, TGI--texture + geometry + image loss) to determine the optimal training objective. We found that using the texture + geometry + image loss produces the best results.}
	\centering
	\begin{tabular}{ p{0.10\textwidth} p{0.12\textwidth} p{0.12\textwidth}  }
    %\begin{tabular}{ lccccc }
		\toprule
		& \multicolumn{2}{c}{Loss Type} \\
		\cmidrule(l){2-3}
        & TG & TGI \\
		\midrule
        Subject 1 & 39.12 & \textbf{33.66} \\
		\midrule
        Subject 2 & 87.67 & \textbf{75.48} \\
		\bottomrule
	\end{tabular}
\label{table:evalloss}
\end{table}

% Table
Table \ref{table:evalloss} shows the quantitative results from these experiments. For each combination of training loss (TG--texture + geometry, TGI--texture + geometry + image) we give the image-based loss on set of 8 validation viewpoints. Incorporating the image-based loss certainly helps lower the MSE although qualitatively the results are similar. We also found that training was unstable without the geometry loss. This is likely because there is not enough long-range signal in the image gradients to effectively estimate geometry.

\begin{figure}[t]
    \centering
    %\subfloat[synthetic to real]{
    %\begin{tabular}{c c}    \includegraphics[width=0.22\textwidth]{figs/trackswitch/switch_id0_01_0_000014.png} & \includegraphics[width=0.22\textwidth]{figs/trackswitch/switch_id0_01_0_000054.png} \\
    %\end{tabular}
    %\label{subfig:switcha}
    %}
    
    %\subfloat[real to synthetic]{
    %\begin{tabular}{c c}
    %\includegraphics[width=0.22\textwidth]{figs/trackswitch/switch_id0_10_0_000027.png} & \includegraphics[width=0.22\textwidth]{figs/trackswitch/switch_id0_10_0_000067.png} \\
    %\end{tabular}
    %\label{subfig:switchb}
    %}
    \includegraphics[trim={0 3mm 0mm 0},width=1\linewidth]{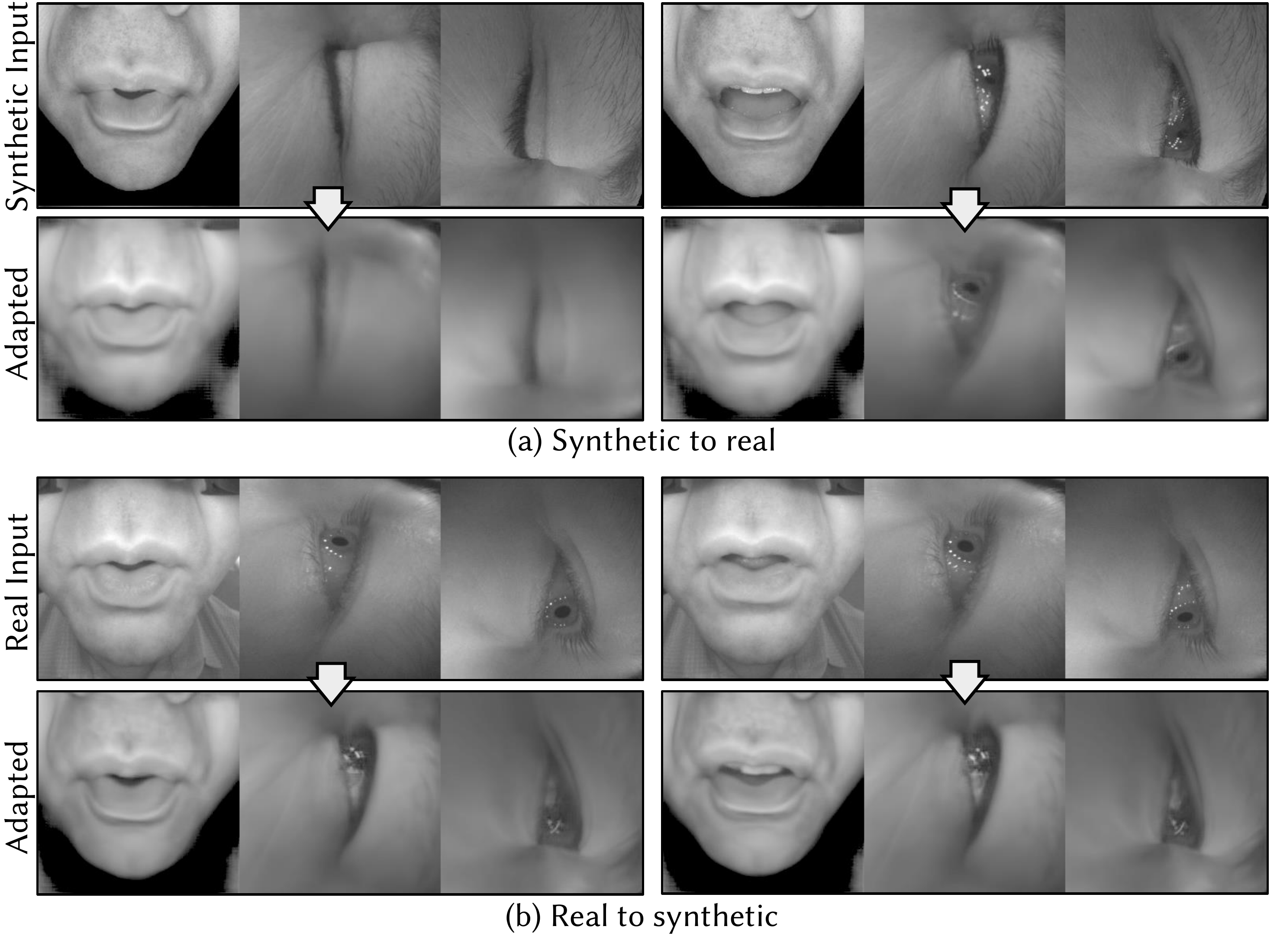}
    \caption{Translating between real headset images to synthetic headset images. Sub-figure (a) shows synthetic headset images (top row) translated to real headset images (bottom row) by switch the conditioning variable in the VAE. Sub-figure (b) shows real headset images (top row) translated to synthetic headset images (bottom row) with the same method). This shows that the $\mathbf{y}$ contains a common representation of facial state regardless of being synthetic or real.}
    \label{fig:vae_translate}
\end{figure}

\begin{figure*}[t]
    \centering
    \includegraphics[width=1.0\textwidth]{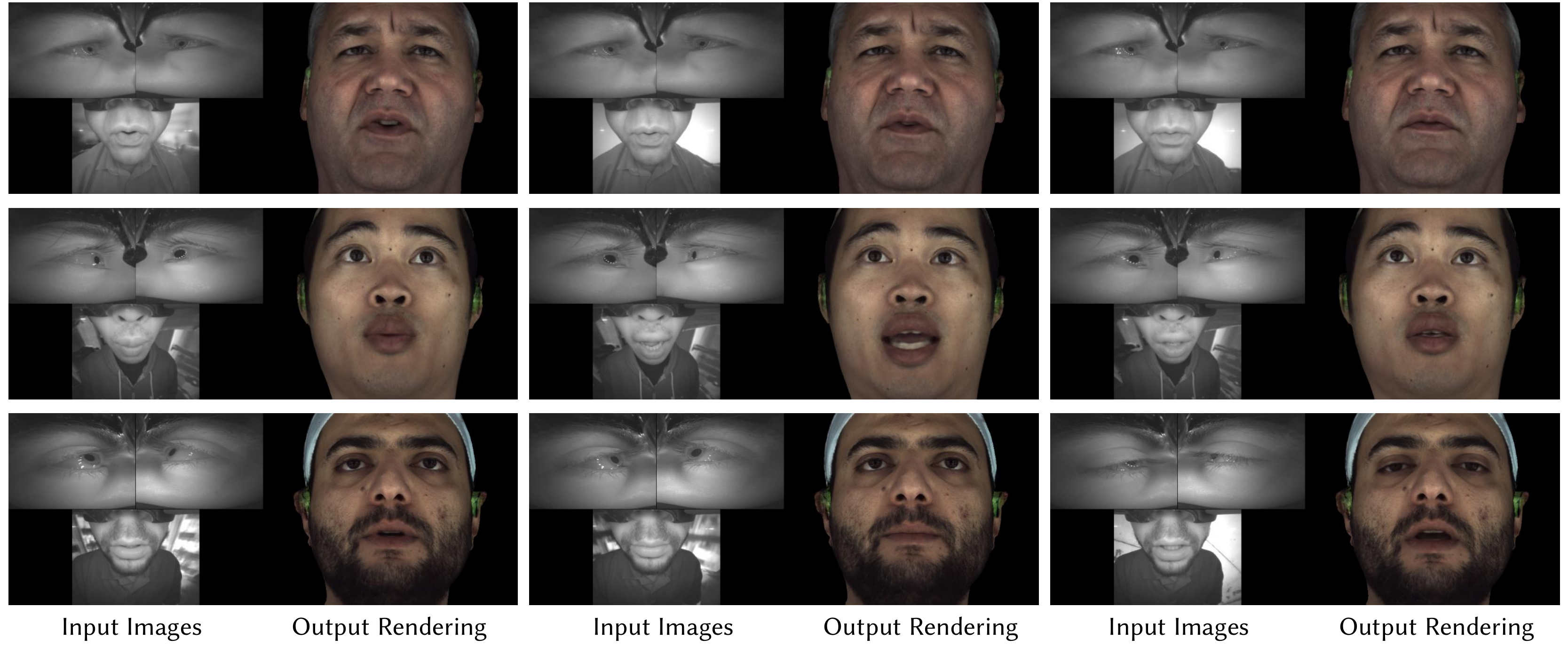}
    %\subfloat[Subject 1]{
    %    \includegraphics[width=0.32\textwidth]{figs/tracking/concat000015.png}
    %    \includegraphics[width=0.32\textwidth]{figs/tracking/concat000028.png}
    %    \includegraphics[width=0.32\textwidth]{figs/tracking/concat000032.png}
    %    %\includegraphics[width=0.24\textwidth]{figs/tracking/concat000040.png}
    %}
    
    %\subfloat[Subject 2]{
    %    \includegraphics[width=0.32\textwidth]{figs/tracking/002576793/concat000031.png}
    %    \includegraphics[width=0.32\textwidth]{figs/tracking/002576793/concat000058.png}
    %    \includegraphics[width=0.32\textwidth]{figs/tracking/002576793/concat000071.png}
    %    %\includegraphics[width=0.24\textwidth]{figs/tracking/002576793/concat000101.png}
    %}
    
    %\subfloat[Subject 3]{
    %    \includegraphics[width=0.32\textwidth]{figs/tracking/002608483/concat000000.png}
    %    \includegraphics[width=0.32\textwidth]{figs/tracking/002608483/concat000014.png}
    %    \includegraphics[width=0.32\textwidth]{figs/tracking/002608483/concat000060.png}
    %    %\includegraphics[width=0.24\textwidth]{figs/tracking/002576793/concat000101.png}
    %}
    \caption{Video-driven Animation. This figure shows three examples per subject of HMD images (left) encoded into the latent code $\mathbf{z}$ and decoded into geometry and appearance and rendered (right). Despite only seeing parts of the face, our animation method captures and reproduces facial state well.}
    \label{fig:video_driven}
\end{figure*}

\subsection{Video-driven Animation}

% Intro
In this section, we show how our tracking VAE ($\mathcal{E}, \mathcal{D}$) builds a common representation of facial state across two different modalities and then we give qualitative results showing our full live animation pipeline. Note that in this work our pipeline is entirely person-specific.

% figure showing decoding cross modality and that it preserves semantics for the most part
Figure \ref{fig:vae_translate} shows examples of encoding with one modality and decoding with the other. We can see in these examples that semantics tend to be preserved when the modality is changed at inference time despite only encoding and decoding the same modality at train time. This is primarily due to the variational autoencoder prior, which encourages the latent space to take on a unit Gaussian distribution.

% TODO: experimental setup
%To achieve video-driven animation, we retarget a generic blendshape-based tracker to our model. The benefit of this is that it can work on any new user. For this experiment, we perform video-based tracking on a set of novel subjects.

% Figure
Figure \ref{fig:video_driven} shows qualitative results of a user driving an avatar in real time. Our system works well not only for expressions but also for subtle motion during speech. The benefit of our model is that it encodes facial state with high accuracy, encoding it into a joint geometry and appearance model. This allows us to model complex changes in appearance due to changes in blood flow, complex materials, and incorrect geometry estimates. Please see our supplemental video for additional examples.

%\subsection{Audio-driven Animation}
%
%% TODO: experimental setup
%
%\begin{figure}[t]
%    \begin{center}
%        \framebox[0.49\textwidth]{\parbox[c][8em][s]{0.49\textwidth}{}}
%    \end{center}
%    \caption{Speech-conditioned Rendering. The top row shows frames from the original recording. The bottom row shows renderings of our model when conditioned on a new speaker saying the same phrase. We can see that the expression are accurate in general while many of the subtitles of motion are preserved.}
%    \label{fig:audio_driven}
%\end{figure}
%
%% Figure: speech conditioning results
%Figure \ref{fig:audio_driven} shows the results of performing speech-conditioned rendering with our model. We can see that the speech-conditioned rendering of the original phrase looks very similar to the original recording. This system allows us to drive any of our avatars with any speech.

\section{Discussion}
\label{sec:Discussion}

In this paper, we presented a method for capturing and encoding human facial appearance and rendering it in real-time. The method unifies the concepts of Active Appearance Models, view-dependent rendering, and deep networks. We showed that the model enables photo-realistic rendering by leveraging multiview image data and that we can drive the model with performance. We believe there are exciting opportunities for extending our approach in different ways.

Our approach is unique because we directly predict a shaded appearance texture with a learned function. This is in contrast to traditional approaches that predict physically-inspired lighting model parameters (e.g., albedo maps, specular maps) which enable relighting. A limitation of our approach in its current form is a limited ability to relight. With improvements to our capture apparatus to allow for dynamic lighting during capture, we believe we can alleviate these limitations by introducing lighting as a conditioning variable.

As our experiments showed, we are able to accurately predict the appearance of the face when our tracked geometry closely matches the true surface of the face. This approach is even effective for regions of the face that are typically very difficult to accurately model because of their complex reflectance properties (e.g., eyes and teeth). Some artifacts still remain, however (e.g., blurring on the teeth) especially where there is no true smooth surface (e.g., hair). We may be able to turn to alternative representations of geometry (e.g., point clouds) to alleviate some of these problems.

Our tracking model enables high-fidelity real-time tracking from cameras mounted on a virtual reality headset by automatic unsupervised correspondence between headset images and multi-camera capture images. In this work, we limited the scope to building person-specific trackers and rendering models. To increase the scalability of our work, we need to modify our approach to support tracking and rendering for arbitrary people. This is a difficult problem because it requires learning semantic facial correspondence between different people.

Finally, we believe that building realistic avatars of the entire body with our approach will help enable self- and social- presence in virtual reality. This task comes with a new set of problems: handling the dynamics of clothing, the difficulties of articulating limbs, and modeling the appearance of interactions between individuals. We are confident that these are tractable problems along this line of research.

\bibliographystyle{ACM-Reference-Format}
\bibliography{facedecoder}

\end{document}